\documentclass[useAMS,usenatbib]{mn2e}
\usepackage{times,txfonts,units,graphicx,aas_macros,natbib,xspace,color}
\def\icarus{{Icarus}}

\newcommand\cm{\,\rm cm}
\newcommand\m{\,\rm m}
\newcommand\s{\,\rm s}
\newcommand\ms{\,\rm m\,s^{-1}}
\newcommand\g{\,\rm g}

\newcommand\K{\,\rm K}
\newcommand\yr{\,\rm yr}
\newcommand\Myr{\,\rm Myr}

\newcommand\km{\,\rm km}

\newcommand\au{\,\rm au}

\newcommand\perccm{\,\rm cm^{-3}}
\newcommand\Msun{\,\rm M_\odot}

\newcommand\yes{$\circ$}
\newcommand\no{$-$}

\newcommand\tms{\!\times\!}
\newcommand\cdt{\!\cdot\!}

\newcommand\xx{\hat{{\mathbf x}}}
\newcommand\yy{\hat{{\mathbf y}}}
\newcommand\zz{\hat{{\mathbf z}}}

\newcommand\V{\mathbf v}
\newcommand\B{\mathbf B}

\newcommand\tauc{\tau_{\rm c}}

\newcommand{\rms}[1]{\left<\right.\!#1\!\left.\right>}
\newcommand{\ee}[1]{$\,\times 10^{#1}$}

\newcommand\Rm{\mathrm{Rm}}

\newcommand{\simgt}%
           {\,\hbox{\lower0.35ex\hbox{$\sim$}\llap{\raise0.35ex\hbox{$>$}}}\,}
\newcommand{\simlt}%
           {\,\hbox{\lower0.35ex\hbox{$\sim$}\llap{\raise0.35ex\hbox{$<$}}}\,}

\newcommand\NIII{\textsc{nirvana-iii}\xspace}

\title[Dynamics of planetesimals in dead zones]%
      {On the dynamics of planetesimals
        embedded in turbulent protoplanetary discs with dead zones}
      \author[Gressel, Nelson \& Turner]
             { Oliver~Gressel$^{1\star}$,
               Richard~P.~Nelson$^{1\star}$ and 
               Neal~J.~Turner$^2$\thanks{E-mail:~
               \{o.gressel,~r.p.nelson\}@qmul.ac.uk; neal.turner@jpl.nasa.gov}\\
               $^1$Astronomy Unit, Queen Mary, University of London,
               Mile End Road, London E1 4NS, United Kingdom\\ 
               $^2$Jet Propulsion Laboratory, California Institute of
               Technology, Pasadena, CA 91109, USA
             }

\begin{document}

\date{Accepted 2011 April 19. %
      Received 2011 April 19; %
      in original form 2011 February 28}

\pagerange{\pageref{firstpage}--\pageref{lastpage}} \pubyear{2002}

\maketitle

\label{firstpage}


\begin{abstract}
Accretion in protoplanetary discs is thought to be driven by
magnetohydrodynamic (MHD) turbulence via the magnetorotational
instability (MRI). Recent work has shown that a planetesimal swarm
embedded in a fully turbulent disc is subject to strong excitation of
the velocity dispersion, leading to collisional destruction of bodies
with radii $R_{\rm p} < 100\km$. Significant diffusion of planetesimal
semimajor axes also arises, leading to large-scale spreading of the
planetesimal population throughout the inner regions of the
protoplanetary disc, in apparent contradiction of constraints provided
by the distribution of asteroids within the asteroid belt.
In this paper, we examine the dynamics of planetesimals embedded in
vertically stratified turbulent discs, with and without dead zones.
Our main aims are to examine the turbulent excitation of the velocity
dispersion, and the radial diffusion, of planetesimals in these discs.
We employ three dimensional MHD simulations using the shearing box
approximation, along with an equilibrium chemistry model that is used
to calculate the ionisation fraction of the disc gas as a function of
time and position. Ionisation is assumed to arise because of stellar
X-rays, galactic cosmic rays and radioactive nuclei.
In agreement with our previous study, we find that planetesimals in
fully turbulent discs develop large random velocities that will lead
to collisional destruction/erosion for bodies with sizes below
$100\km$, and undergo radial diffusion on a scale $\sim 2.5\au$ over a
$5\Myr$ disc life time. But planetesimals in a dead zone experience a
much reduced excitation of their random velocities, and equilibrium
velocity dispersions lie between the disruption thresholds for weak
and strong aggregates for sizes $R_{\rm p} \le 100\km$.  We also find
that radial diffusion occurs over a much reduced length scale $\sim
0.25\au$ over the disc life time, this being consistent with solar
system constraints.
We conclude that planetesimal growth via mutual collisions between
smaller bodies cannot occur in a fully turbulent disc.  By contrast, a
dead zone may provide a safe haven in which km-sized planetesimals can
avoid mutual destruction through collisions.

\end{abstract}



\section{Introduction}

The formation of planetesimals is a key stage in the formation of
planetary systems, but at present there is little consensus about the
processes involved. One class of models envisages an incremental
process in which small dust grains collide, stick and settle
vertically within a protoplanetary disc, with continued growth through
coagulation eventually forming km-sized planetesimals
\citep{1993prpl.conf.1031W, 2000SSRv...92..295W}.  Models discussed by
\citet{2000SSRv...92..295W}, that assume the disc is laminar and has a
surface density a few times larger than the minimum mass solar nebula
model of \citet{1981PThPS..70...35H}, predict planetesimal formation
times that are a few thousand local orbital periods at each location
in the disc, implying formation times of $\approx$ few $\times
10^4\yr$ at $5\au$. More recent models presented by
\citet{2008A&A...480..859B}, that examine planetesimal formation in
low density turbulent discs, conclude that incremental planetesimal
formation is prevented by the rapid inward migration of metre-sized
boulders when they form, combined with fragmentation induced by high
velocity collisions caused by differential radial migration.  Large
km-sized bodies were only obtained in these models when
unrealistically large values for the critical velocity for
fragmentation were adopted, and a significant increase in the
dust-to-gas ratio above solar abundance was assumed. A local
particle-in-a-box model of growth through dust coagulation presented
by \citet{2010A&A...513A..57Z}, that incorporates recent experimental
results on the compactification of porous aggregates, suggests that
coagulative growth beyond millimetre sizes is difficult to achieve
because compacted aggregates tend to bounce rather than stick. These
latter results raise serious questions about the validity of the
incremental picture of planetesimal formation.

A number of the difficulties faced by the incremental model have been
known for many years, leading to searches for alternative planetesimal
formation scenarios. Gravitational instability of the dense dust layer
that forms through vertical settling was suggested by
\citet{1973ApJ...183.1051G} as a means of forming km-sized
planetesimals. Although this pathway is not now widely accepted,
recent models of planetesimal formation that involve collective
effects, including self-gravity of regions of enhanced solids
concentration, have been proposed. \citet{2008ApJ...687.1432C} have
developed a model in which mm-sized chondrules are concentrated in
turbulent eddies, and then contract on the settling time of the solid
grains under the action of self-gravity to form large $100\km$-sized
planetesimals.  The combined action of the streaming instability
\citep{2005ApJ...620..459Y}, and trapping of metre-sized boulders in
short-lived turbulent eddies, has been suggested as a means of
inducing planetesimal formation via gravitational collapse by
\citet{2007Natur.448.1022J}, leading to the direct formation of
planetesimals that are as large as the asteroid Ceres ($R_{\rm p}
\simeq 500 \km$). Although these methods of forming planetesimals
avoid the problem of growing beyond the metre-sized barrier, they each
have their own difficulties, and as yet no clear consensus of how
planetesimals form has emerged.

Once planetesimals have formed, runaway growth may lead to the
formation of $\sim 1000\km$ oligarchs on time scales of $\sim 10^5\yr$
\citep{1993Icar..106..190W,2009ApJ...690L.140K}, which then accumulate
smaller bodies to become planetary embryos and cores during the
oligarchic phase \citep{1993Icar..106..210I,
1998Icar..131..171K}. Rapid runaway growth of km-sized planetesimals
requires that their velocity dispersion remains significantly smaller
than the escape velocity of the accreting bodies, which for
$10\km$-sized objects is $\sim 10\ms$. A self-stirring swarm of
planetesimals embedded in a laminar disc can maintain a low velocity
dispersion via gas drag and dynamical friction, but the
situation in a turbulent disc may be very different due to stochastic
forcing of planetesimal random motions by the turbulence. Planetesimal
accretion that occurs at a rate determined by the geometric cross
section may lead to planetary formation time scales in excess of
observed disc life times of a few Myr \citep{2001ApJ...553L.153H}.

A further constraint is that planetesimal collision velocities must
remain below the catastrophic disruption threshold for collisional
growth to occur. For planetesimals between the sizes 1 - $10\km$,
disruption thresholds lie approximately in the interval $10\ms \le
v_{\rm crit} \le 60\ms$, depending on the material strength and
whether the planetesimals are monolithic bodies or rubble piles
\citep{1999Icar..142....5B,2009ApJ...691L.133S}. Clearly an external
source of stirring, such as disc turbulence, may not only prevent
runaway growth, but may lead to the collisional destruction of
planetesimals.

Turbulence is generally believed to provide the anomalous source of
viscosity required to explain the typical $\sim 10^{-8}{\rm
M}_{\odot}\yr^{-1}$ accretion rates of T Tauri stars
\citep{2004AJ....128..805S}.  The most promising source of this is MHD
turbulence driven by the magnetorotational instability
\citep[MRI,][]{1991ApJ...376..214B}. Both local shearing box and global
simulations indicate that the non linear, saturated state of the MRI
is vigorous turbulence that is able to transport angular momentum at a
rate that can explain observations of T Tauri accretion rates
\citep{1995ApJ...440..742H,1998ApJ...501L.189A,2001ApJ...554..534H,
2003MNRAS.339..983P}. But the MRI requires that the gas is
sufficiently ionised in order for the linear instability to operate
\citep{1994ApJ...421..163B}, and for non linear turbulence to be
sustained \citep{2000ApJ...530..464F}.  In the absence of sufficient
free electrons in the gas phase, the flow returns to a laminar state.

Protoplanetary discs surrounding T Tauri stars are cold and dense,
leading to low levels of ionisation \citep{1983PThPh..69..480U}.  In
the main body of these discs the primary sources of ionisation are
expected to be external: stellar X-rays \citep{1997ApJ...480..344G}
and possibly galactic cosmic rays \citep{1981PASJ...33..617U}, each of
which have a limited penetration depth into the disc.
\citet{1996ApJ...457..355G} presented a model in which the disc
surface layers are ionised, and sustain a turbulent accretion flow,
whereas the disc interior is a `dead zone' where the flow remains
laminar and minimal accretion takes place. This layered accretion disc
model has been the subject of numerous investigations over recent
years which have examined the role of dust grains
\citep{2000ApJ...543..486S, 2006A&A...445..205I}, gas-phase heavy
metals \citep{2002MNRAS.329...18F}, different chemical reaction
networks \citep{2006A&A...445..205I}, the role of turbulent mixing
\citep{2006A&A...445..223I}, and the Hall effect
\citep{1999MNRAS.307..849W}.

MHD simulations have also been used to examine the dynamics of
magnetised discs in which a finite electrical conductivity plays an
important role. \citet{2000ApJ...530..464F} examined the saturated
level of turbulence as a function of the magnetic Reynolds
number. \citet{2003ApJ...585..908F} examined disc models in which
magnetic resistivity varied with height, simulated the layered
accretion model proposed by \citet{1996ApJ...457..355G}, and
demonstrated that the dead zone maintains a modest Reynolds stress due
to waves being excited by the active layers.  More recent studies have
coupled time dependent chemical networks to the MHD evolution using a
multi-fluid approach, and have demonstrated that turbulent diffusion
of charge carriers into the dead zone can enliven it, in the absence
of small dust grains \citep{2007ApJ...659..729T,
2008A&A...483..815I}. The presence of a significant population of
small grains, however, reduces the time for removal of free charges
from the gas phase, and the dead zone is maintained
\citep{2008ApJ...679L.131T}.

Low mass planets and planetesimals embedded in turbulent
protoplanetary discs are subject to stochastic gravitational forces
due to turbulent density fluctuations
\citep{2004MNRAS.350..849N,2005A&A...443.1067N}.  The stochastic
forcing leads to diffusion of the semimajor axes and excitation of the
eccentricity of the embedded bodies. \citet{2010MNRAS.409..639N} --
hereafter paper~I -- examined the dynamics of planetesimals embedded
in fully turbulent cylindrical disc models without dead zones, and
demonstrated that the excitation of the velocity dispersion will lead
to collisional disruption of planetesimals of size $R_{\rm p} \le
10\km$. They also demonstrated that very good agreement can be
obtained between local shearing box simulations and global simulations
when shearing boxes of sufficient size are employed \citep[see
also][]{2009ApJ...707.1233Y}. In this paper, we extend the work
presented in paper~I and examine the dynamics of planetesimals
embedded in vertically stratified discs, with and without dead zones,
using shearing box simulations. A key result is that the turbulent
stirring of planetesimals is significantly reduced in discs with dead
zones, possibly allowing for their continued growth rather than
destruction or erosion during collisions. As such, we propose that
dead zones may provide safe havens for planetesimals, and are a
required ingredient for the formation of planetary systems.

This paper is organised as follows: We describe the physical model and
numerical methods in Section~\ref{sec:methods}. The presentation of
the results is divided into four parts: In Section~\ref{sec:gas_dyn},
we describe the general morphology and dynamics of the emerging
hydromagnetic turbulence. Section~\ref{sec:trq} discusses how the
resulting stochastic gravitational torques experienced by bodies
embedded in the disc can be classified by their distribution and
auto-correlation. Finally, in Sections~\ref{sec:e-growth} and
\ref{sec:da-grow}, we analyse the temporal evolution in eccentricity
and radial diffusion of a swarm of particles immersed in the flow. We
discuss the implications of our results for planetary formation in
Section~\ref{sec:discussion}, and we draw our conclusions in
Section~\ref{sec:conclusion}.


\section{Methods}
\label{sec:methods}

For the models reported-on in this paper, we make use of the
second-order Godunov code \NIII
\citep{2004JCoPh.196..393Z,2008CoPhC.179..227Z} and solve the standard
equations of resistive magneto-hydro\-dynamics (MHD) in a local,
corotating Cartesian frame ($\xx,\yy,\zz$), which read:
\begin{eqnarray}
  \partial_t\rho +\nabla\cdt(\rho\V) & \!=\! & 0\,, \nonumber\\\ 
  \partial_t(\rho\V) +\nabla\cdt
          [\rho\mathbf{vv}+p^\star\!\!-\mathbf{BB}] & \!=\! &
           2\rho\Omega\,( q \Omega x\,\xx - \zz\tms\V )
           - \rho\nabla\Phi(z) \zz  \nonumber\\ 
  \partial_t \B -\nabla\tms(\V\tms\B -\eta \nabla\tms\B) & \!=\! &  0\,,
  \nonumber\\      \nabla\cdt\B & \!=\! &  0\,,\nonumber
\end{eqnarray}
with a static gravitational potential $\Phi(z) =
\frac{1}{2}\,\Omega^2\,z^2$, the total pressure $p^{\star} = p +
\frac{1}{2}\B^2$, and where all other symbols have their usual
meanings. The first source term in the momentum equation includes the
Coriolis acceleration and tidal force in the shearing box
approximation \citep[see][for implementation
details]{2007CoPhC.176..652G,2010ApJS..189..142S}. For the shear rate,
$q$, we apply the Keplerian value of $-3/2$.

\subsection{Model geometry and parameters}

The simulations are semi-global in the sense that we only consider a
small horizontal patch of a protoplanetary accretion disc but include
the full vertical structure. Our fiducial stratified simulation covers
$\pm4$ pressure scale heights, $H$, to account for hydrostatic
stratification under the assumption of an isothermal equation of
state. To accommodate sufficiently wide MRI-active regions in the
presence of a realistic dead zone, we use an increased box size of
$5.5$ scale heights on each side, i.e., a box with $L_z=11\,H$. This
provides an extra $3$ to $4\,H$ above and below the nominal dead zone,
whose vertical extent either side of the midplane is $\sim 2H$.

Based on our previous work on permissible shearing box sizes for
studying the dynamics of embedded bodies in turbulent discs in
paper~I, we chose a standard horizontal extent of $4\times16\,H$. Due
to the higher computational cost of the models including a dead zone,
we adopt a somewhat smaller horizontal domain size of $3\times12\,H$,
yet still large enough to allow for the excitation of spiral density
waves \citep{2009MNRAS.397...52H, 2009MNRAS.397...64H}.  The box sizes
and grid resolution for the different models are listed in
Tab.~\ref{tab:models}.

Because the ionisation model used to compute the resistivity involves
chemical rate equations, we need to specify a unit system to convert
between computational units and physical units for the mass density
and temperature. To allow for direct comparison with other results, we
scale our model such that it approximates the widely used
``minimum-mass'' protosolar nebula model
\citep{1981PThPS..70...35H}. More specifically, we place our local box
at a radius $r_0=5\au$ and choose a disc aspect ratio $h \equiv H/r =
0.05$, a column density of $\Sigma=135\g\cm^{-2}$, a temperature
$T=108\K$, and an isothermal sound speed of $c_s=667\ms$.
Using our definition of $H$, the equilibrium vertical density
profile is given by 
$\rho(z) =\rho_0 \exp{\left[-z^2/(2 H^2)\right]}$,
where $\rho_0$ is the midplane density.

\begin{table}\begin{center}
\begin{tabular}{lccccc}\hline
& domain $\ [H]$& resolution & XR & SR & CR \\[4pt]
\hline
A1        & $4\tms16,\,\pm 4.0$& $128\tms256\tms256$
          & \no & \no & \no \\
D1        & $3\tms12,\,\pm 5.5$& $\ 72\tms144\tms264$
          & \yes & $\times 10$ & \yes \\
D2        & $3\tms12,\,\pm 5.5$& $\ 72\tms144\tms264$
          & $\times 20$ & $\times 10$ & \yes \\ 
B1        & $4\tms16\tms2$& $128\tms512\tms64$
          & \no & \no & \no \\ 
\hline
\end{tabular}
\end{center}
\caption{Simulation parameters for the studied models. Columns 4-6
  indicate the ionisation flux for X-rays (XR), short-lived
  radionuclides (SR), and cosmic rays (CR), respectively; open circles
  indicate nominal values, and a multiplication symbol followed
  by an integer gives the multiple of the nominal value adopted.}
  \label{tab:models}
\end{table}

\subsection{Initial and boundary conditions}

For the initial magnetic field, we apply the standard zero-net-flux
magnetic configuration $B_z(x)\sim\sin(2\pi\,x/L_x)$, with an initial
plasma parameter (ratio of thermal to magnetic pressure) $\beta_{\rm
p}\simeq50$. To avoid low $\beta_{\rm p}$ in the far halo region, we
attenuate $B_z(z)$ with height, as to keep the plasma parameter
roughly constant. At the same time, to obey the divergence constraint,
we introduce a radial field $B_x(x,z)$, which is subsequently sheared
out into an azimuthal field as the system evolves. The described
configuration has proven to produce a relatively smooth transition
into developed turbulence, avoiding extreme field topologies in the
linear growth phase.

Following \citet{2008ApJ...679L.131T}, we furthermore impose a weak
additional $B_z$ net field, with an associated midplane $\beta_{\rm p}=13900$.
Due to the assumed periodicity, the flux
associated with this field is preserved for the duration of the
simulations, and we found this to be a useful means of maintaining
active MRI over long evolution times in the presence of a dead
zone. Such a field may be a local residual of a large-scale field
dragged in by the disc during its formation from the collapse of a
magnetised molecular cloud. We also note that in the presence of a
mean-field dynamo mechanism \citep[as e.g. discussed
in][]{2010MNRAS.405...41G}, one would expect a significant net
vertical flux in any given local patch of the disc.

For the velocity field, we use vertical boundary conditions which
allow material to leave the box but prevent inflow. This has the
advantage that the magnetic field can escape the box rather than
pile-up near the domain boundary, as is the case for a completely
closed box \citep{1996ApJ...463..656S}. The magnetic field boundary
conditions implement a rough approximation to an external vacuum field
and are usually referred to as ``pseudo-vacuum'' conditions. This
means that the horizontal components are set to zero on the boundary,
and only a vertical field component (with vanishing gradient) is
allowed. We note that this type of boundary enforces a gradient in
$B_x$ and $B_y$ near the top and bottom of the domain, which results
in an accelerated wind in this region, clearing material and embedded
field structures from the magnetically dominated halo region. To
compensate for the mass loss associated with the outflow boundaries
and to enforce a stationary background, we continually re-instate the
initial density profile in each grid cell by means of an artificial
mass source term \citep[cf.][]{2009A&A...498..335H}.

In the simulations adopting a dead zone due to insufficient ionisation
we prescribe a locally and temporally varying magnetic diffusivity
$\eta$.  Some earlier simulations of dead zones by
\citet{2003ApJ...585..908F} and \citet{2007ApJ...670..805O} prescribed
a static diffusivity profile derived from a physically motivated
ionisation model.  Here we further extend the realism of the
diffusivity model by deriving a time-variable profile based on the gas
column density, the detailed procedure of which is described in the
following section.

After our disc models have achieved a steady turbulent state, we
introduce 25 planetesimals into the disc midplanes.  We treat these
planetesimals as non interacting test particles which evolve under the
full 3D influence of the gravitational field of the turbulent
disc. The disc gravity is softened on a length scale equal to the cell
diagonal.

\subsection{The diffusivity model}

Comparing different chemical reaction networks,
\citet{2000ApJ...543..486S} and \citet{2006A&A...445..205I} found that
the adsorption of free electrons and ions onto the surfaces of small
dust grains plays an important role in removing free charges from the
gas phase. These authors find that even tiny mass fractions of
micron-sized grains can significantly deplete charge carriers.

Taking this dominance of small grains into account, we here refrain
from following the detailed non-equilibrium chemistry in combination
with a multi-fluid treatment as was done by
\citet{2007ApJ...659..729T} and \citet{2008A&A...483..815I}. Instead
we adopt a simplified treatment of the gas-phase reactions. As a
reasonable approximation to the more intricate modelling outlined
above, we assume that recombination happens much faster than any
dynamical mixing timescale in our system, which seems warranted in the
presence of small grains \citep{2000ApJ...543..486S,
2006A&A...445..205I, 2006A&A...445..223I}. As a first step towards
more realistic description, we allow the magnetic diffusivity,
$\eta=\eta({\bf x},t)$, to vary spatially in all three dimensions, and
update it according to a look-up table derived from the reaction
network in \texttt{model4} of \citet{2006A&A...445..205I}. In
particular, we assume dust grains of size $0.1\mu{\rm m}$ and with
density $3\g\perccm$ and use a dust-to-gas mass ratio of
$10^{-3}$. The assumed gas-phase Magnesium abundance is taken to be
depleted by a factor $10^{-4}$ compared to its solar value. The input
parameters to the tabulated diffusivity are: (i) mass density, (ii)
gas temperature, and (iii) the local ionisation rate $\zeta$. All
three quantities are evaluated on a per-grid-cell basis. To include
the effects of external irradiation on $\zeta$, we compute column
densities to both the upper and lower disc surfaces.

\subsubsection{Ionisation sources}

There are several likely sources of ionising radiation in the vicinity
of a newly born star. \citet{2009ApJ...703.2152T} have recently
studied how contributions from stellar X-rays, radionuclides, and
energetic protons (interstellar cosmic rays, protons accelerated in
the coronae of the star and disc) influence the shape and extent of a
possible dead zone. We have implemented their ionisation model,
including all ionisation sources. Because of uncertainties related to
some of the proposed mechanisms, we take a conservative standpoint in
this paper and focus on stellar X-ray irradiation (XR), and
interstellar cosmic rays (CRs) as the prime sources of ionisation.

Observations of young stars in Orion show that T Tauri star X-ray
luminosities vary by approximately four orders of magnitude, with a
median value of $L_{\rm XR} \simeq 2 \times 10^{30}\,{\rm
erg\,s}^{-1}$ \citep{2000AJ....120.1426G}. \citet{1999ApJ...518..848I}
performed Monte-Carlo radiative transfer calculations of the
ionisation rate in a standard protostellar disc model, adopting the
above value for $L_{\rm XR}$ and assuming a plasma temperature of
$k\,T = 5\,{\rm keV}$. Applying a fit to their results, we approximate
the ionisation rate due to X-rays by
\begin{equation}
  \zeta_{\rm XR} = 2.6\tms10^{-15}\s^{-1}\ 
  \left[ {\rm e}^{-\Sigma_a/\Sigma_{\rm XR}}
       + {\rm e}^{-\Sigma_b/\Sigma_{\rm XR}}
  \right]\,r_{\au}^{-2}\,,
\end{equation}
where $r_{\au}$ is the position of our box in astronomical units,
$\Sigma_a$ and $\Sigma_b$, are the gas column densities above and
below a given point, and $\Sigma_{\rm XR}=8.0\g\cm^{-2}$ is the
absorption depth for the assumed spectrum of X-rays. As detailed in
\citet{2009ApJ...690...69U}, we prescribe the following vertical
attenuation of interstellar CRs illuminating the disc surfaces:
\begin{equation}
\zeta_{\rm CR} = 5\tms10^{-18}\s^{-1}
  \ {\rm e}^{-\Sigma_a/\Sigma_{\rm CR}}\ 
  \left[ 1+\left(\frac{\Sigma_a}{\Sigma_{\rm CR}}\right)^\frac{3}{4}
  \right]^{-\frac{4}{3}}
  +\ \dots
\end{equation}
Here, $\Sigma_{\rm CR}=96\g\cm^{-2}$ is the cosmic ray attenuation
depth estimated by \citet{1981PASJ...33..617U}, and the dots indicate
the corresponding contribution from the second column density
$\Sigma_b$.

\begin{figure}
  \includegraphics[height=0.62\columnwidth]{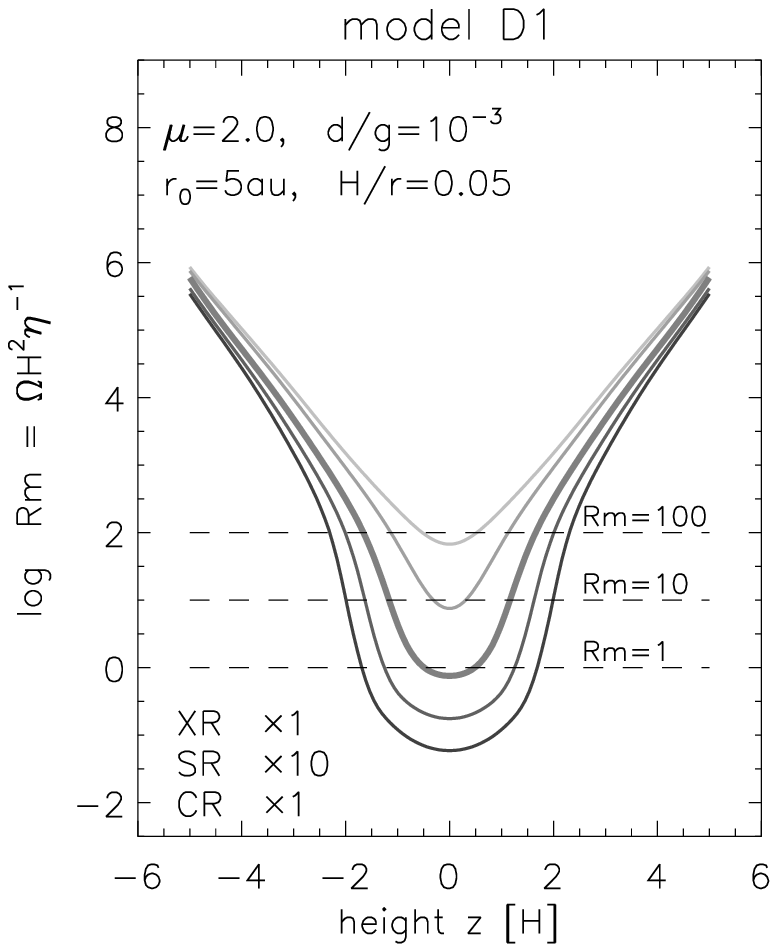}\hfill%
  \includegraphics[height=0.62\columnwidth]{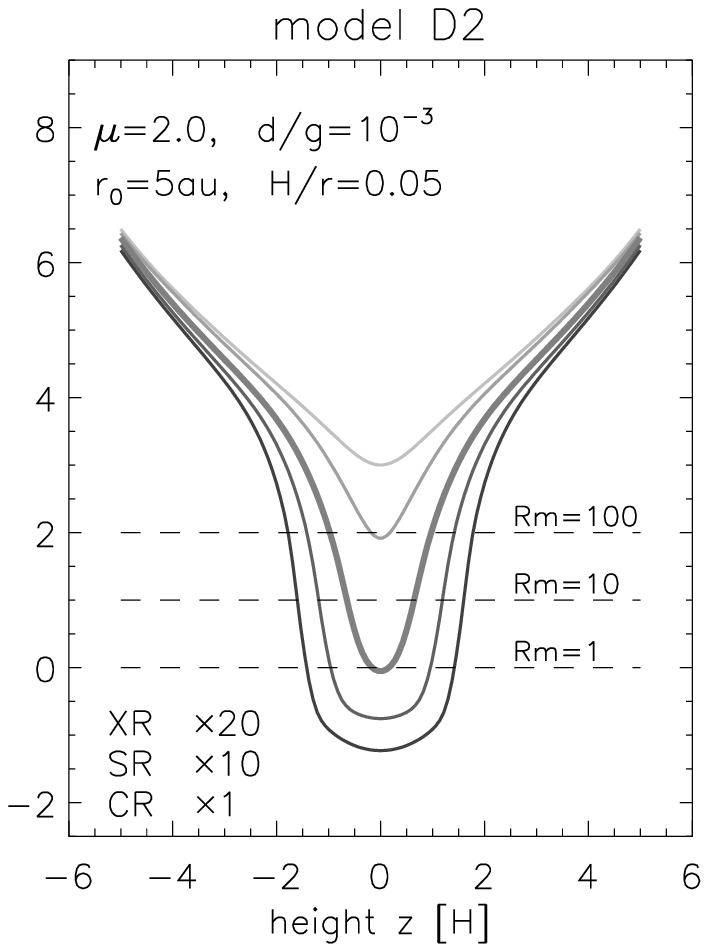}
  \caption{Vertical $\eta$~profiles for runs D1 and D2, in terms of
    the magnetic Reynolds number, $\Rm=\Omega\,H^2\,\eta(z)^{-1}$, as
    given by our ionisation model. The thick line corresponds to the
    adopted initial model (with $\Sigma=135\g\cm^{-2}$). Thin lines on
    either side are for disc masses increased and reduced by factors
    of two, and demonstrate the strong dependence on the mass
    column. A dead zone is expected for $\Rm\simlt10$
    \citep[cf.][]{2008ApJ...679L.131T}.}
  \label{fig:ppd_strat}
\end{figure}

Because the numerically allowed diffusive time step decreases with the
grid spacing squared, adequately resolved dynamical simulations can
become prohibitive in the presence of high diffusion
coefficients. Given limited computational resources, we therefore
decided to confine the dynamical span in the magnetic diffusivity
$\eta$ to a reasonable range. We primarily do this by including an
ambient ionisation due to the decay of short-lived radionuclides (SR),
which we further enhance by a factor of ten compared to the fiducial
value of $\zeta_{\rm SR}= 3.7\tms10^{-19}\s^{-1}$ given in
\citet{2009ApJ...703.2152T}. As can be seen in
Fig.~\ref{fig:ppd_strat}, where we plot the resulting $\eta$~profiles,
the floor in the magnetic Reynolds number due to the ambient level of
ionisation is well below the expected threshold for the formation of a
dead zone. This implies that we should not expect our results to be
significantly affected by enhancing the effects of decaying
radionuclides.

We remark that, compared to previous work, our diffusivity profiles
overall imply lower magnetic Reynolds numbers. Given that
\citet{2000ApJ...530..464F} have found a critical $\Rm_{\rm c}\simeq
10^4$ for sustained MRI in a zero-net-flux configuration, we expect
that the amplitude of an external net-flux will have a significant
impact on the level of turbulence.

\subsubsection{Sub-cycling}

Notwithstanding the moderate range in $\eta$, the numerical constraint
given by the diffusive propagation of information is one of the main
limiting factors in our simulations. To avoid a degradation of the
numerical accuracy in the MRI-active region by an excessively low
numerical time step in the dead zone, we choose to account for the two
separate regimes in $\Rm$. This is done by operator-splitting the
diffusive term in the induction equation and applying a sub
cycling-scheme for its update\footnote{We typically chose a
sub-cycling ratio of $5-8$ to approximately match the restriction
given by the high Alfv\'{e}n speeds in the halo.}. By doing so, we are
able to integrate the non-diffusive part of the MHD equations with a
longer time step, enhancing the accuracy of the solution in regions of
low magnetic diffusivity.

\subsection{Code improvements}

Properly resolving the growing modes of the magneto-rotational
instability (MRI) with Godunov-type codes has been found to depend on
the reconstruction strategy used and on the ability of the Riemann
solver to capture the Alfv\'{e}nic mode \citep{2010arXiv1003.0018B}. To
improve the representation of discontinuities in our finite volume
scheme, we recently extended the \NIII code with the Harten--Lax--van
Leer Discontinuities (HLLD) approximate Riemann solver introduced by
\citet{2005JCoPh.208..315M}. To guarantee the required directional
biasing of the electromotive force interpolation
\citep[cf.][]{2010A&A...516A..26F}, we have implemented and tested the
upwind reconstruction procedure of \citet{2005JCoPh.205..509G}.

To enhance the stability of our code in the strongly magnetised
corona, we gradually degrade the reconstruction order from second- to
first-order accuracy near the vertical domain boundaries, thus
avoiding undershoots in the hydrodynamic state variables in strong
shocks.

\subsubsection{Artificial mass diffusion}

To facilitate the study of a low-beta disc corona,
\citet{2000ApJ...534..398M} have introduced the concept of a so-called
Alfv{\'e}n speed limiter, circumventing prohibitively high signal
speeds in low-density regions. Such a limiter can in principle be
adopted for the approximate Riemann solvers we use. We chose, however,
a different approach and instead add an artificial mass diffusion term
to the equations of mass and momentum conservation. The diffusion
coefficient is chosen to lie well below the truncation error in the
bulk of the domain. In grid cells where the density contrast exceeds a
specified dynamic range, $C_{\rm dyn}$, the coefficient is gradually
adjusted according to
\begin{equation}
  D(\rho) = \frac{1}{6}\ \left[1+\left(10^{C_{\rm dyn}}\, 
                           \frac{\rho}{\rho_0}\right)^4\right]^{-1/4}
\end{equation}
such that the grid Reynolds number approaches order unity --
coinciding with the stability limit for the explicit time integration
of the diffusive term. The transition function is chosen in a way that
the inverse operation can be efficiently implemented by means of
consecutive square-root operations. We typically chose $C_{\rm
dyn}=5$, resulting in four to five orders of magnitude in density
contrast. We note that, because the diffusive fluxes are part of the
finite volume update, this approach does not violate mass
conservation, and therefore avoids the problems of enforcing an
artificial floor value in the density. In conclusion, our approach has
proven to greatly benefit the overall robustness without noticeably
affecting the solution in the interior of the domain. Similarly to the
limiters used by \citet{2000ApJ...534..398M} and
\citet{2008A&A...490..501J}, the Courant-Friedrichs-Lewy time-step
constraint due to the fast magnetosonic mode is significantly
alleviated, allowing for economical use of computing resources.


\section{Disc model properties}
\label{sec:gas_dyn}

\begin{table*}
  \begin{minipage}[b]{1.3\columnwidth}
\begin{tabular}{lcccccc}\hline
& orbits & $\rms{\alpha}$ & $\sigma_{\Gamma}\ \ [\cm^2\s^{\!-2}]$ & 
$\tauc\ \ [2\pi\Omega^{-1}]$ & $C_{\sigma}(\Delta x)\ \ [H]$ &
$C_{\sigma}(e)\ \ [H/r]$ \\[4pt]
\hline
A1     & 217 & 0.0105 & 0.45\ee{8} & 0.30 & 5.21\ee{-3} & 2.68\ee{-3}\\
D1     & 224 & 0.0038 & 0.06\ee{8} & 0.29 & 4.72\ee{-4} & 1.54\ee{-4}\\
D2     & 223 & 0.0051 & 0.13\ee{8} & 0.27 & 7.25\ee{-4} & 2.50\ee{-4}\\
B1$^*$ & 505 & $0.05\quad$ & 0.74\ee{8} & 0.32 & 7.70\ee{-3} & 2.77\ee{-3}\\
\hline
\end{tabular}\\
$^*$torque corrected for 2D/3D evaluation \citep[cf. Fig.~8
  in][]{2010MNRAS.409..639N}.
\end{minipage}
\begin{minipage}[b]{0.7\columnwidth}
\caption{Overview of simulation results. Gravitational torques
  $\Gamma$ represent the standard deviation $\sigma$ of a normal
  distribution (see Fig.~\ref{fig:trq_his}), Coherence times $\tauc$
  are from a fit to the torque ACF (see Fig.~\ref{fig:trq_acf}). The
  random walk amplitudes $C_\sigma(\Delta x)$ and $C_{\sigma}(e)$
  refer to the dispersion in the displacement and eccentricity of a
  swarm of particles. }
\label{tab:sim_results}
\end{minipage}
\end{table*}

\begin{figure}
  \center\includegraphics[width=0.85\columnwidth]{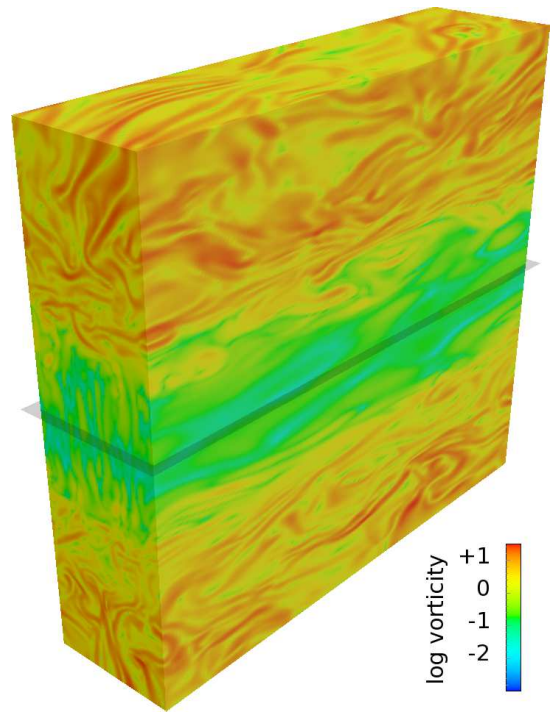}
  \caption{Visualisation of the turbulent flow structure. The colour
    coding indicates the vorticity, $\log_{10}\
    |\nabla\times\mathbf{u}|$ of the perturbed flow
    $\mathbf{u}=\V-q\Omega x\yy$. While the turbulent regions above
    and below the dead zone show strongly folded vortex structures,
    the flow pattern near the midplane is dominated by shearing
    waves. The contrast in the vorticity amplitude is $\sim100$.}
  \label{fig:vort}
\end{figure}

In the following, we will present results from three different
simulations (see Tab.~\ref{tab:models} for details). To be able to
make a quantitative comparison with respect to the effects of a dead
zone, we performed a fiducial stratified model with fully active MRI,
referred to as model A1. In general, this model agrees well with the
unstratified simulations presented in paper~I. For our standard dead
zone model, D1, we choose the reference ionisation rates of
\citet{2009ApJ...703.2152T} along with a dust-to-gas mass ratio of
1$\,:\,$1000, i.e., accounting for a modest depletion of the smallest
grains by coagulation and sedimentation. The resulting diffusivity
profile is plotted in the left panel of Figure~\ref{fig:ppd_strat}.

The resulting dead zone in this model covers roughly $\pm2H$, making
it a reasonable proxy for what a realistic dead zone at $5\au$ might
look like. Gaseous nebulae around newly forming stars, however, are
subject to substantial variations in X-ray irradiation
\citep{2000AJ....120.1426G}, so for our second dead zone model, D2, we
increase the stellar flux by a factor of twenty. We find the average
total thickness of the dead zone to be reduced by roughly $1.5$
pressure scale heights in this case (see Fig.~\ref{fig:ppd_strat},
right panel).

\subsection{Hydromagnetic turbulence}

One important result of the early simulations of layered accretion
discs by \citet{2003ApJ...585..908F} was that the dead zone, despite
its name, retains a non-negligible level of Reynolds stresses, namely
in the form of waves that are excited in the active layers. These
residual motions can clearly be seen in Fig.~\ref{fig:vort}, where we
have visualised the flow structure in terms of the logarithmic
vorticity. The colour coding exhibits very distinct patterns in the
two regions: while the MRI active layers show folded vortex features
characteristic of developed turbulence, the dead zone region is
clearly non-turbulent, and dominated by sheared density waves with
$k_R\gg k_\phi$.  At the interface between the two zones, one can
furthermore see structures indicating some level of turbulent mixing
into the dead zone. Because our model assumes that recombination on
the surface of small dust grains is efficient (leading to a
recombination time scale much smaller than the dynamical mixing time
scale) this does not, however, have an effect on the extent of the
dead zone.

\begin{figure}
  \center\includegraphics[width=\columnwidth]{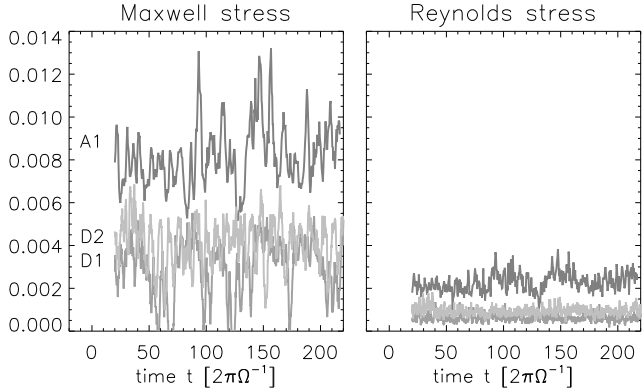}
  \caption{Comparison between the box averaged turbulent Maxwell- and
    Reynolds stresses. Values are normalised by the gas pressure in
    the disc midplane, $p_0$. The time-averaged sums of these, i.e.,
    the $\alpha$ parameters, are listed in column three of
    Tab.~\ref{tab:sim_results}.}
  \label{fig:transp}
\end{figure}

Our primary aim in this paper is to examine statistically the
dynamical evolution of planetesimals embedded in turbulent discs with
and without dead zones. In order to do this we require a
quasi-stationary model of turbulence.  As can be seen in
Fig.~\ref{fig:transp}, where we plot the time variation of the
box-averaged transport coefficients, our models fulfil this
requirement of a stationary mean with (admittedly strong) superposed
fluctuations. Because the bulk of the transport is associated with the
MRI-active regions, the turbulent stresses only seem to depend weakly
on the actual width of the dead zone. We remark that due to the
different field configuration, the overall strength of the turbulence
is somewhat weaker than in our previous non-stratified models
presented in paper~I.  This is related to buoyant loss of magnetic
flux from the disc in stratified simulations, and is consistent with
recent results in the literature \citep[cf.][and references
therein]{2010ApJ...713...52D}.

\begin{figure*}
  \center\includegraphics[width=2\columnwidth]{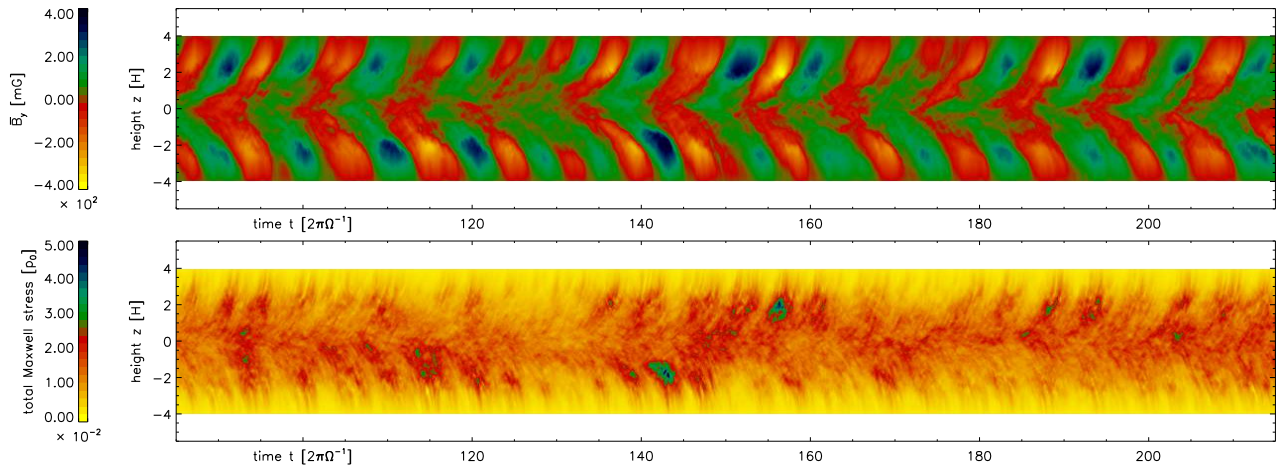}
  \caption{Spatio-temporal evolution of (a) mean toroidal field, (b)
    total Maxwell stress, for the fully active model A1.}
  \label{fig:spctm_A1}
\end{figure*}

\subsection{Spatio-temporal disc evolution}

Figure~\ref{fig:spctm_A1} shows space-time plots of the mean toroidal
magnetic field $\bar{B}_y(z,t)$, as well as the total Maxwell stress
for the fully active model A1. Probably the most striking features in
this plot are the periodic ``butterfly'' patterns in the large-scale
magnetic field, which are commonly observed in this type of simulation
\citep[e.g.][]{2010ApJ...713...52D,2010MNRAS.tmp.1450F}. Because the
upwelling field structures are not associated with a bulk motion of
the flow, the likely explanation for the emerging patterns is a
mean-field dynamo as recently re-investigated by
\citet{2010MNRAS.405...41G}. Even though we see quite strong
fluctuations in the total Maxwell stress, the overall state is quasi
stationary. Note that the two most prominent peaks of the total
Maxwell stress (i.e. around 140-145 and 155-160 orbits) are (a) due to
a strong mean field, and (b) confined to one half of the box. The
situation of magnetic activity being dominant in one half of the disc
has been observed before \citep[see e.g. fig.~7
in][]{2000ApJ...534..398M}, and might be explained by equally strong
quadrupolar and dipolar contributions. These cancel on one side and,
at the same time, add-up on the opposite side, resulting in the
observed lopsided appearance. In fact, we cannot define a clear parity
of the field in Fig.~\ref{fig:spctm_A1}, which implies that the
dipolar and quadrupolar dynamo modes seem to possess similar growth
rates.

\begin{figure*}
  \center\includegraphics[width=2\columnwidth]{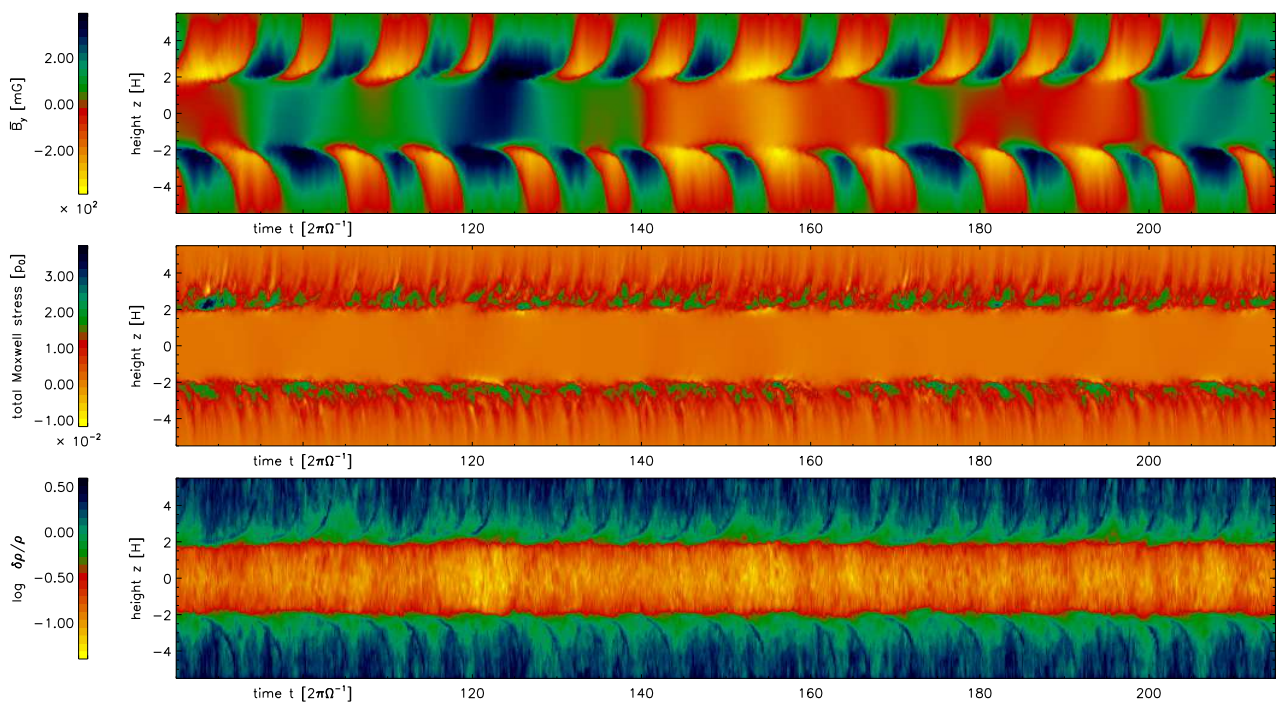}
  \caption{Same as Fig.~\ref{fig:spctm_A1}, but for the dead zone
    model D1, and including an extra panel showing the relative
    density fluctuations (logarithmic). The horizontal field exhibits
    the same dynamo cycles in the MRI active layers. As already seen
    by \citet{2008ApJ...679L.131T}, the toroidal field leaks into the
    diffusively dominated midplane region, contributing to the
    stresses. Despite the absence of strong turbulent fields, density
    fluctuations reach a level of 10-20\% within the dead zone.}
  \label{fig:spctm_D1}
\end{figure*}

In the following discussion, we will focus on the vertical structure
of the disc and how it is shaped by the presence of a dead zone. For
this we will show the time-averaged\footnote{If not stated otherwise,
we average over the interval [20,210] orbits.} vertical profiles of
various quantities. To demonstrate that this is a meaningful
procedure, we provide space-time diagrams of three representative
quantities (Fig.~\ref{fig:spctm_D1}) for model D1. Again, the mean
toroidal field generally is of irregular parity, i.e., neither
quadrupolar nor dipolar symmetry is seen to prevail. Whether the
concept of a global parity is relevant in the presence of a highly
diffusive ``insulating'' layer remains a matter of discussion. It is
interesting to note, however, that MRI channel modes -- which are a
non-linear solution to local net-flux MRI simulations -- represent
global modes even when stratification is included
\citep{2010MNRAS.406..848L}. In fact, we see some indication of
channels in our simulations, most prominently in the density
\citep[compare the lower panel in Fig.~\ref{fig:spctm_D1}
with][fig.~3]{2010MNRAS.406..848L}.

Quite remarkably, the number of field reversals related to the
dynamo-cycle seems to be largely unaffected by the presence of the
dead zone. As mentioned above, one main motivation in explaining the
rising patterns by a mean-field dynamo, was the absence of a bulk
motion near the midplane -- implying that the pattern speed cannot be
explained by the Parker instability \citep{2010ApJ...708.1716S}. The
key in explaining the \emph{upward} motion was a \emph{negative}
buoyant $\alpha$~effect near the midplane
\citep{2000A&A...362..756R,2008AN....329..725B}. With the pattern in
the disc halo being unchanged in the presence of a dead zone, this
picture possibly has to be amended. Naively, for a positive
$\alpha$~effect in the halo \citep[as found
by][]{2010MNRAS.405...41G}, one would not expect an outward butterfly
diagram from classical theory -- demonstrating the need for non-linear
feedback both in the diagnostics \citep{2010A&A...520A..28R} as well
as in the modelling \citep*{2009MNRAS.398.1414B,2010AN....331..667C}.

We note however, that because of the presence of the weak net vertical
flux in our current simulations, they are not strictly comparable to
our previous work. Due to the net flux, a possible explanation of the
unchanged migration direction might be the presence of prevailing
channel modes.  These were shown to produce a strong negative
helicity \citep[cf. fig.~4 in][]{2010MNRAS.405...41G}, compatible with
the upward motion of the wave patterns\footnote{This was already
suggested as an alternative scenario in the mentioned paper.}.

Now focusing our attention to the actual dead zone region, we see that
a significant amount of azimuthal (and in fact, radial) field diffuses
in to the low-$\Rm$ region around the midplane. This leakage has first
been seen in simulations by \citet{2007ApJ...659..729T} and
\citet{2008ApJ...679L.131T}, which are very similar to ours and also
contain a weak net vertical field. In contrast to this, the
simulations by \citet{2003ApJ...585..908F},
\citet*{2007ApJ...670..805O} and \citet{2008A&A...483..815I}, which
apply a zero-net-flux configuration, do not show such an effect. It
seems peculiar, however, that the presence of a net flux should play a
role in this context. As \citet{2008A&A...483..815I} illustrate in
their fig.~2, the assumed diffusivity profiles of
\citet{2003ApJ...585..908F} result in moderately high magnetic
Reynolds numbers near the midplane. On the contrary, in the model of
\citet*{2007ApJ...659..729T} and also in our model
(cf. Fig~\ref{fig:ppd_strat}) the magnetic Reynolds number reaches
much lower values of $\Rm<10$ for $|z|<H$, leading to a much reduced
diffusion time scale, possibly facilitating the leakage of large-scale
field from the disc halo into the dead zone.

Finally, we note that relative density fluctuations remain at a level
of up to ten per cent within the dead zone -- as seen in the lowermost
panel of Fig.~\ref{fig:spctm_D1}. Episodes of lower perturbance (as
indicated by lighter colours) are clearly correlated with strong
azimuthal field, and notably with a negative Maxwell stress at the
interface between the dead and active zones. As already mentioned
above, we see indications of persistent channel modes (dark streaks
above $|z|\simgt 2H$), which are not unexpected given the magnetic
configuration. It might be argued that such features are overly
pronounced at the current numerical resolution, and might be destroyed
by parasitic instabilities
\citep{2009ApJ...698L..72P,2009MNRAS.394..715L,2010MNRAS.406..848L}
when the resolution is further increased.

\begin{figure}
  \center\includegraphics[width=0.9\columnwidth]{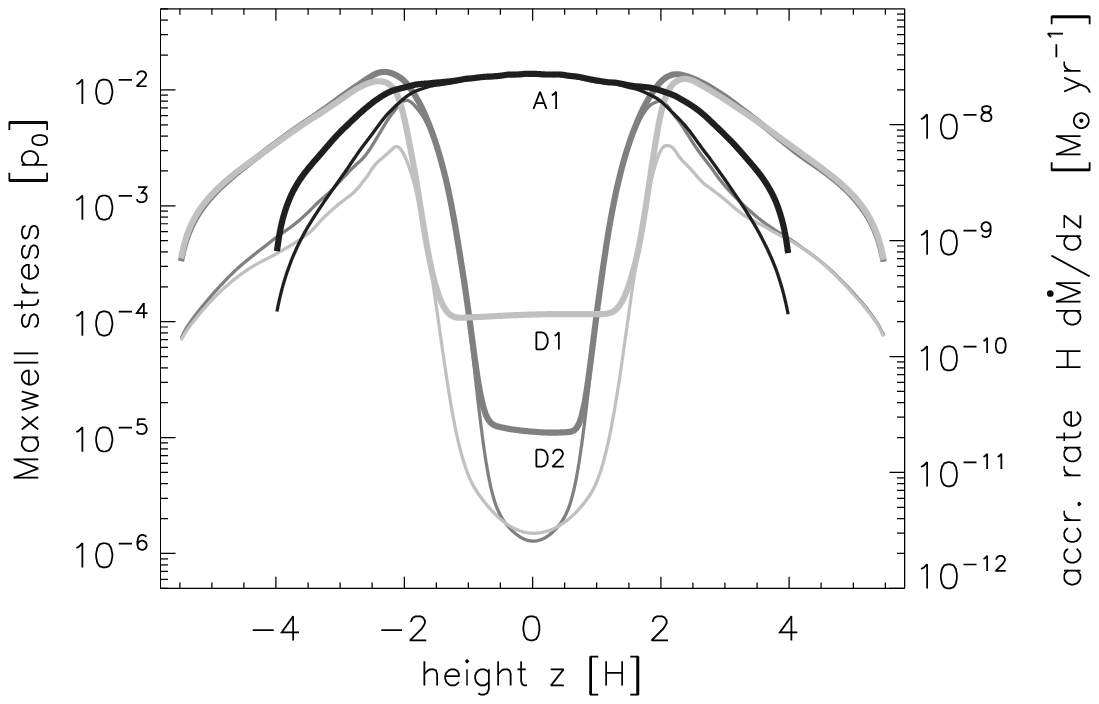}
  \caption{Maxwell stresses for the different models. Thick lines
    indicate the total stress $\propto -\rms{B_R\,B_\phi}$, thin lines
    represent the turbulent contribution $\propto -\rms{B'_R\,
    B'_{\phi}}$, where $B'\!=\!B-\rms{B}$. The vertically integrated
    mass accretion rates (see rhs axis) for the models D2, D1 and A1,
    are $9.85\times10^{-8}$, $7.82\times10^{-8}$, and
    $1.36\times10^{-7}\Msun\yr^{-1}$, respectively. Note the unexpected
    trend in the depth between models D1 and D2, as opposed to
    Fig.~\ref{fig:Reyn_z}.}
  \label{fig:Maxw_z}
\end{figure}

\subsection{Disc vertical structure}

Figure~\ref{fig:Maxw_z} shows time-averaged vertical profiles of the
$R\phi$~components of the Maxwell stress tensor. Because we use an
isothermal equation of state, resulting in a constant sound speed,
$c_{\rm s}$, these profiles translate directly into a fractional mass
accretion rate as indicated by the right-hand axis. The values on this
axis are given by the expression $H\,{\rm d}\dot{M}/{\rm
d}z=3\pi\,c_s\,H^2\,\rho(z) T_{r\phi}/p(z)$, which can be derived from
the usual approximation for the mass accretion rate in a steady disc
$\dot{M} = 3 \pi \nu \Sigma$, but relaxing the assumption that the
effective viscous stresses are independent of $z$.

The vertically integrated mass flow rates generated by the magnetic
stresses are $9.85 \times 10^{-8}$ and $7.82 \times 10^{-8} \Msun
\yr^{-1}$ for runs D2 and D1, respectively. This similarity is
expected because the heights of the dead and active zones 
about the midplane in each model are similar ($\sim
1.5 H$ and $2 H$, respectively, for the dead zones
in models D2 and D1). The mass flow
rate in the fully active run A1 is $1.36\times10^{-7}\Msun\yr^{-1}$,
nearly twice that observed in the model with the largest dead
zone. Given that the bulk of the mass in the disc lies within the dead
zone, where the Maxwell stresses for models D2 and D1 are factors of
$\sim 10^{-2}$ and $10^{-3}$ smaller than for model A1, the bulk of
the transport in models D1 and D2 happens in the active layers because
the stresses (and effective viscosity coefficient $\alpha$) remain
large there.

Figure~\ref{fig:Maxw_z} shows that the Maxwell stresses in the
midplane regions are dominated by large-scale fields in runs D1 and D2
-- even more so than in the halo regions of the fully active run
A1. This result agrees broadly with recent simulations
\citep[cf. fig.~4 in][]{2008ApJ...679L.131T} which included similar
physical effects. While the turbulent contribution to the Maxwell
stress (thin lines) falls-off to $\simeq10^{-6}$ in the dead zone for
both models D1 and D2, one can clearly see that the effect of the
greater dead zone width in model D1 is compensated for by a shallower
valley in the total Maxwell stress (thick lines). The inverse top-hat
shape of the profiles is likely related to large-scale field diffusing
into regions of intermediate $\Rm$, as seen in the topmost panel of
Figure~\ref{fig:spctm_D1}. Given that the transition region is closer
to the low-$\beta$ halo for model D1 (with a wider dead zone), it
seems plausible that the strength of the regular field is larger in
this case. Clearly, the contribution due to the fluctuating fields
becomes negligible in both cases.

\begin{figure}
  \center\includegraphics[width=0.9\columnwidth]{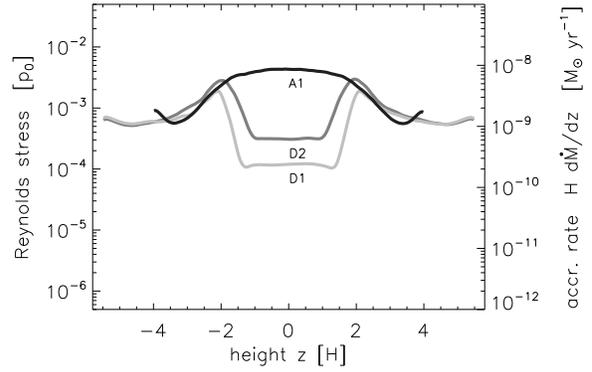}
  \caption{Same as Fig.~\ref{fig:Maxw_z}, but for the Reynolds stress
  $\rms{\rho\,v_R\delta v_\phi}$. The vertically-integrated mass
  accretion rates are $2.08\times10^{-8}$, $1.39\times10^{-8}$, and
  $3.64\times10^{-8}\Msun\yr^{-1}$ for models D2, D1, and A1,
  respectively.}
  \label{fig:Reyn_z}
\end{figure}

Considering that, for model D1, the floor value of the Maxwell stress
is of the same magnitude as the Reynolds stress
(cf. Fig.~\ref{fig:Reyn_z}), such large-scale stresses might play an
important role in mediating mass transport within the dead zone --
making a more detailed future study of the discussed trend worthwhile,
particularly in the context of global disc models, where it may be
possible to observe large scale accretion flows in dead zones.

In Figure~\ref{fig:Reyn_z}, we plot vertical profiles of the
time-averaged Reynolds stress. As expected, the lines agree very well
in the MRI-active regions. Comparing the models D1 and D2, we now see
a correspondence between the dead zone width and depth, as expected.
It appears that larger and more massive active zones lying above and 
below the dead zone are more able to excite higher amplitude sound 
waves which are able to propagate into the dead zone, generating a 
correspondingly larger Reynolds stress there.

Based on two runs with different (zero net flux) field topologies,
\citet{2003ApJ...585..908F} concluded that ``the Reynolds stress in
the midplane does not seem to depend on the size of the dead zone but
rather the amplitude of the turbulence in the active layers''. This is
clearly not the case for our models D1 and D2, which show quite
different Reynolds stresses in the dead zone, while having almost
identical levels of Maxwell stresses -- both in the total and
fluctuating part -- within the active region (see
Figs.~\ref{fig:Maxw_z} and \ref{fig:Reyn_z}, respectively).  While we
reckon that the larger Reynolds stress is related to the larger mass
of the active layer in model D2, this discrepancy may also be due to
the presence of large-scale fields diffusing into the midplane region
in our simulations. Note that \citet{2003ApJ...585..908F} did not
mention such fields in their discussion.

\section{Characterising the gravitational forcing}
\label{sec:trq}

\begin{figure*}
  \includegraphics[height=0.54\columnwidth]{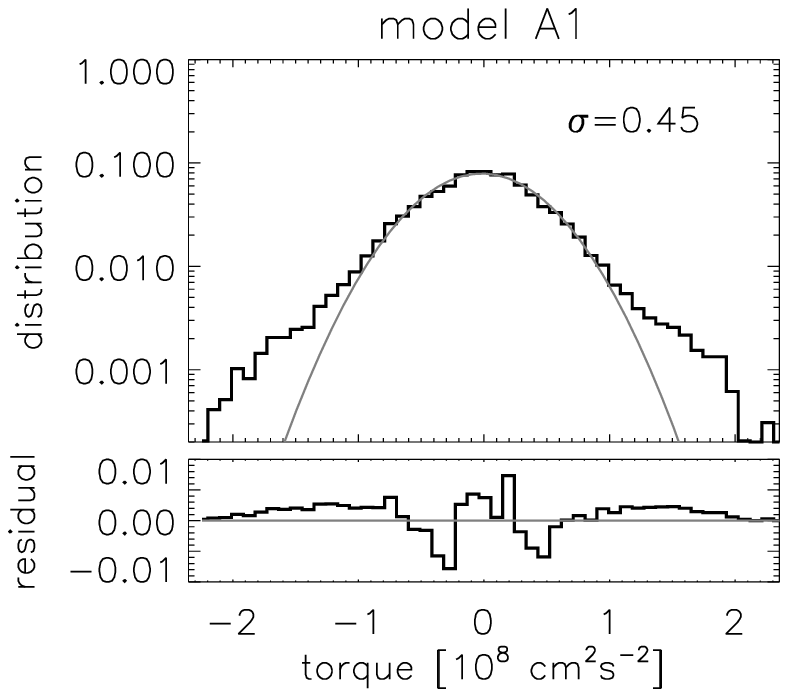}\hspace{2ex}%
  \includegraphics[height=0.54\columnwidth]{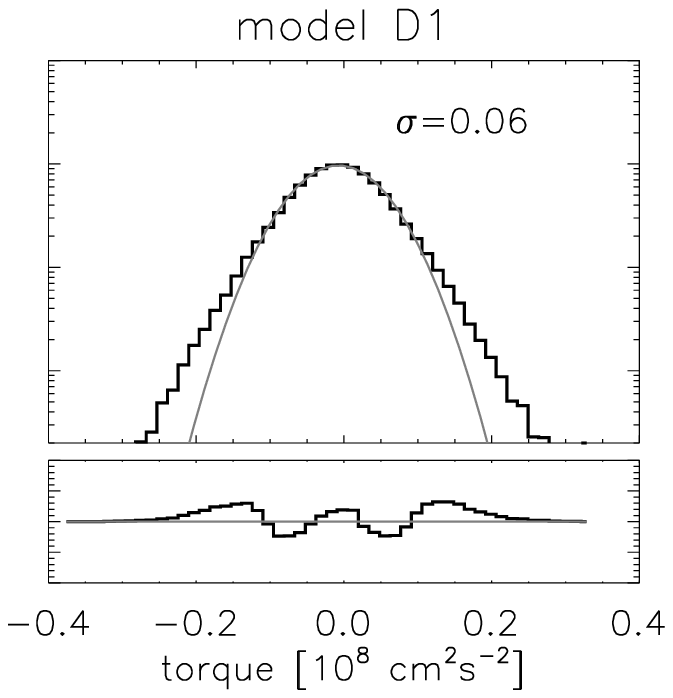}\hspace{2ex}%
  \includegraphics[height=0.54\columnwidth]{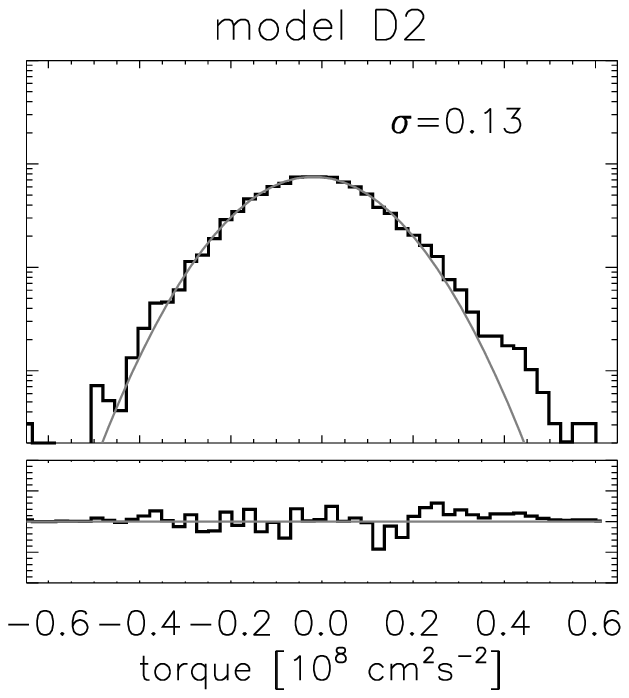}
  \caption{Torque histograms for models A1, D1 and D2. The grey
    lines indicate fitted normal distributions, for which standard
    deviations are given (note the different scale on the
    abscissae). While the torques within the dead zone are well
    approximated by a Gaussian, the distribution in model A1
    develops some deviations for large torques.}
  \label{fig:trq_his}
\end{figure*}

As first suggested by the global simulations of
\citet{2005A&A...443.1067N}, the stochastic gravitational forces
associated with density fluctuations from developed MRI turbulence have
the potential to limit severely the growth of planetesimals.
In paper~I, we have confirmed this original finding in the framework
of global simulations and local simulations with large enough box sizes,
but in which the vertical component of gravity was neglected.
In the following section, we extend this work to the case of stratified 
discs with and without a dead zone. To facilitate a direct comparison, 
we here focus on the case of a moderately strong net vertical field. In 
this respect, the results should be interpreted as upper limits. To obtain
lower limits for the torque amplitude, models with nominal dead zones
and applying a zero-net-flux configuration will be required.

\subsection{Stochastic torques}

Figure~\ref{fig:trq_his} shows distribution functions of gravitational
torques along the $y$-direction in our simulation box. The histograms
are computed from the time series of a set of 25 particles in each
simulation run. Unlike \citet{2007ApJ...670..805O}, we do not observe
transients in the time series, and the torques are indeed
quasi-stationary in the interval [20,210] orbits, which has been used
to obtain each histogram. As required for a simplified treatment in
terms of a stochastic Monte Carlo model, the histograms are well
represented by a normal distribution. Notably, there are moderate
power-law tails in the fully active run A1, indicating some level of
non-Gaussian fluctuations (also cf. fig.~7 in paper~I). The amplitude
of the torques in the different runs roughly scales with the
square-root of the midplane Reynolds stress. This arises because
the density fluctuations scale linearly with the velocity fluctuations,
and coincides with the scaling found in section 3.3 of 
\citet{2010ApJ...709..759B}.

Apart from the torque amplitude, the ability to stochastically amplify
particle motions depends on the degree of coherence in the fluid
patterns. Building on our work from paper~I, we here apply the same
formalism and study temporal torque autocorrelation functions
(ACFs). We use the fitting formula introduced in section 3.3 of
paper~I, i.e.
\begin{equation}
  S_{\Gamma}(\tau) = \left[ (1-a) + a\,\cos(2\pi\,\omega\,\tau) \right]\,
  {\rm e}^{-\tau/\tau_{\rm c}} \label{eq:fit_acf}\,,
\end{equation}
with free parameters, $a$, $\omega$, and $\tauc$. As discussed in
paper~I, we assume two components to better fit the shape of the ACF:
(i) an exponential decay representing the truly stochastic part of the
torque time series; (ii) a modulation due to ``wavelike''
behaviour, i.e., a negative dip in the ACF reflecting the fact that
density enhancements swashing over the region of influence will
produce torques of opposite signs when approaching and receding the
test particle. Such a feature is also seen in the corresponding
Figs.~9 and 15 of \citet{2007ApJ...670..805O} and
\citet{2009ApJ...707.1233Y}, respectively, complicating the
interpretation of the results.

In paper~I, we have discovered certain trends in the amplitude of the
latter effect, which are probably related to the finite size of the
computational domain. Because the periodic box in some respect acts as
a resonator, the wavelike component probably appears too pronounced in
local simulations. This demonstrates that one does well keeping in
mind the limitations of local boxes in terms of the allowed dynamics
\citep[also see][]{2008A&A...481...21R}. The parameter of interest, of
course, is the correlation time $\tauc$ of the \emph{stochastic}
perturbation. While waves represent ordered motion, it is this
stochastic part that ultimately produces the random-walk amplification
of the particle motions.

\begin{figure}
  \includegraphics[height=0.58\columnwidth]{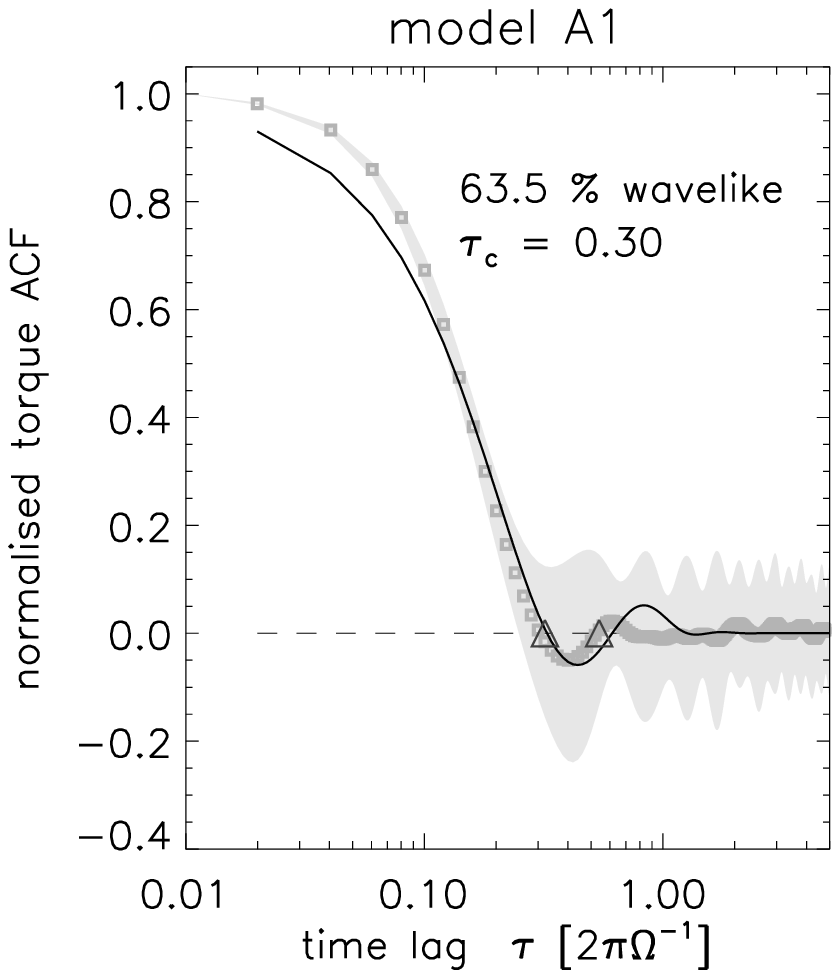}\hfill%
  \includegraphics[height=0.58\columnwidth]{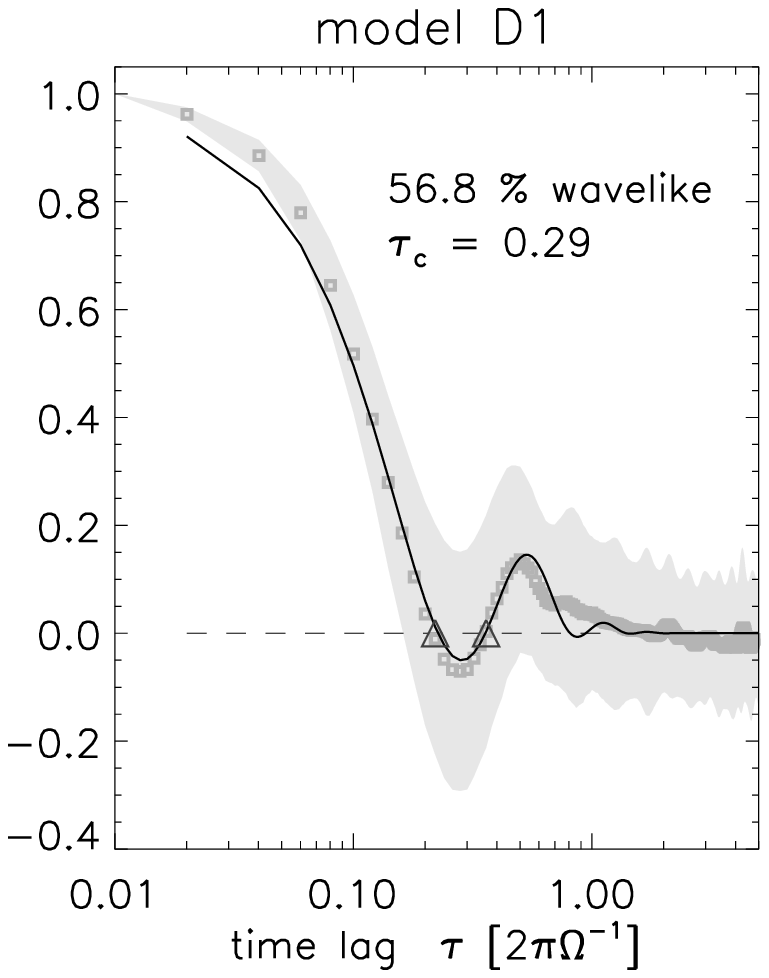}
  \caption{Torque autocorrelation function (ACF) for models A1 and
    D1. The black lines indicate model fits (see text), showing that
    motions are coherent over approximately one third of the orbital
    time scale. The ACF for model D2 is very similar to D1 and is
    hence not shown.}
  \label{fig:trq_acf}
\end{figure}

\begin{figure}
  \begin{minipage}[t]{0.59\columnwidth}
    \includegraphics[width=0.9\columnwidth]{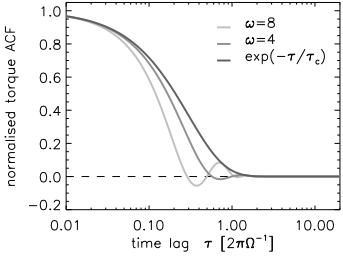}
  \end{minipage}
  \begin{minipage}[b]{0.40\columnwidth}
    \caption{Comparison of three different model ACFs as given by
      Eqn.~(\ref{eq:fit_acf}). All curves have the same coherence time
      $\tauc=0.3$ and yet show quite different zero-crossings.}
    \label{fig:trq_acf_cartoon}
  \end{minipage}
\end{figure}

In the following, we rely on the approach taken in
Eqn.~(\ref{eq:fit_acf}) providing an effective deconvolution of the
two effects. The resulting fits for models A1 and D1 are shown in
Fig.~\ref{fig:trq_acf}. Unlike model B1 (cf. fig.~9 in paper~I), model
A1 shows a moderate level of modulation. Given the horizontal box
dimensions of $4\tms16H$, this is somewhat unexpected as we observed a
consistent trend towards weaker modulation for larger box sizes in
paper~I. This being said, the fitted coherence time of $\tauc=0.30$ in
model A1 is in striking agreement with our earlier unstratified
model. The torque ACF of the dead zone model D1 shows a comparable
amplitude in the modulation and has an almost identical value of
$\tauc=0.29$.

With respect to the artificial forcing function used in the 2D global
simulations of \citet{2010ApJ...709..759B}, we note that our value for
$\tauc=0.3$ coincides with the one stated for their ``reduced
lifetime'' case. Comparing our ACF with the one in their Fig.~2, there
appears, however, to be a discrepancy of a factor of two in the first
zero crossing. This can possibly be explained by a different
modulation of the exponential decay as exemplified in
Figure~\ref{fig:trq_acf_cartoon}. All curves have a common
$\tauc=0.3$. Moreover, the first two curves have a 60\% wavelike
modulations with frequency $\omega$ as indicated, while the third is a
simple exponential. Owing to the confusion with respect to the
definition of the coherence time, it seems worthwhile to further study
how the spectrum of modes affects the apparent
coherence. Nevertheless, given the actual type of forcing used in
\citet{2010ApJ...709..759B}, i.e. a spectrum of wavelike motions, one
may be surprised how similar the ACFs indeed look. Going back to
Fig.~\ref{fig:vort}, it however becomes obvious that it is not the
turbulence we need to approximate but only the spiral density waves
with uncorrelated phases that it creates. 

The corresponding ACF for model D2 is omitted here, as it is very
similar to D1. The fitted coherence time for model D2 is $\tauc=0.27$,
in good agreement with the other two runs. We therefore conclude that
the presence (and depth) of a dead zone does have little influence on
the temporal torque statistics and merely affects the amplitude of the
stirring process.

\begin{table*}\begin{center}
\begin{tabular}{lccccccc}\hline
Model & size & $v_{\rm disp}$ [$\ms\,$] & $v_{\rm disp}$ [$\ms\,$] &
               $v_{\rm crit}$ [$\ms\,$] & $v_{\rm crit}$ [$\ms\,$] &
               $v_{\rm crit}$ [$\ms\,$] & $t_{\rm eq}$ [$\yr\,$] \\[2pt]
      &      & (gas drag)      & (collisional)               
             & (weak)          & (strong)       & (rubble pile) &      \\[4pt]
\hline
A1  & $100\m$   & 73.8  & 79.9    & 1.0   &  7.7    &  -    & $1.91\tms10^4$ \\
    & $1\km$    & 159.0 & 251.4   & 2.9   &  15.2   &  -    & $8.85\tms10^4$ \\
    & $10\km$   & 342.8 & 795.0   & 26.3  &  137.0  & 58.5  & $4.11\tms10^5$ \\
    & $100\km$  & 738.2 & 2514.3  & 262.6 &  1368.0 & 585.4 & $1.91\tms10^6$ \\
\hline
D1  & $100\m$   & 11.0  & 4.5     & 1.0   &  7.7    &  -    & $2.20\tms10^4$ \\
    & $1\km$    & 23.7  & 14.3    & 2.9   &  15.2   &  -    & $2.20\tms10^5$ \\
    & $10\km$   & 51.0  & 45.4    & 26.3  &  137.0  & 58.5  & $2.21\tms10^6$ \\
    & $100\km$  & 110.0 & 143.5   & 262.6 &  1368.0 & 585.4 & $10.3\tms10^6$ \\
\hline
D2  & $100\m$   & 15.2  & 7.4     & 1.0   &  7.7    &  -    & $2.22\tms10^4$ \\
    & $1\km$    & 32.7  & 23.3    & 2.9   &  15.2   &  -    & $2.21\tms10^5$ \\
    & $10\km$   & 70.5  & 73.9    & 26.3  &  137.0  & 58.5  & $1.60\tms10^6$ \\
    & $100\km$  & 153.4 & 233.6   & 262.6 &  1368.0 & 585.4 & $7.58\tms10^6$ \\
\hline
\end{tabular}
\end{center}
\caption{Results for the equilibrium velocity dispersions
obtained for each model as a function of the planetesimal
sizes, and as a function of the damping mechanism.
Also tabulated is the time required to grow to the smaller of the
equilibrium velocity dispersion values given for each model each
size.}
\label{tab:results}
\end{table*}

\section{Eccentricity stirring and planetesimal accretion}
\label{sec:e-growth}

To study the excitation of the velocity dispersion and the radial
diffusion of embedded boulders, planetesimals, and protoplanets, in
each run we integrated the trajectories of a swarm of 25
non-interacting test particles subject to perturbations from the
gravitational potential of the gas disc. Unlike in paper~I, we here
focus on larger bodies and neglect the effects of aerodynamic
interaction. This is warranted for solids with radii above $\simeq
100\m$ (cf. paper~I where the results for planetesimals with sizes in
the range $100\m$ - $10\km$ were found to be similar for run times on
the order of a few hundred planetesimal orbits). The upper limit for
the size range considered here is given by the constraint that
perturbations of the disc remain weak such that spiral wave excitation
and gap-opening can be ignored. This is the case for planetesimals and
small protoplanets, where feedback onto the disc can be ignored.

Although we do not consider in detail the dynamics of metre-sized
boulders that are strongly coupled to the gas via drag forces in
this paper, we recall from paper~I that such bodies quickly achieve
random velocities very similar to the turbulent velocity field of the
gas. The midplane r.m.s. turbulent velocity for the fully turbulent
model A1 is found to be $v_{\rm turb}=(76.6 \pm6.5)\ms$, and for model
D1 $v_{\rm turb}=(17.8 \pm2.6)\ms$. For model D2 $v_{\rm turb}=(28.5
\pm3.6)\ms$.

\begin{figure}
  \center\includegraphics[width=\columnwidth]{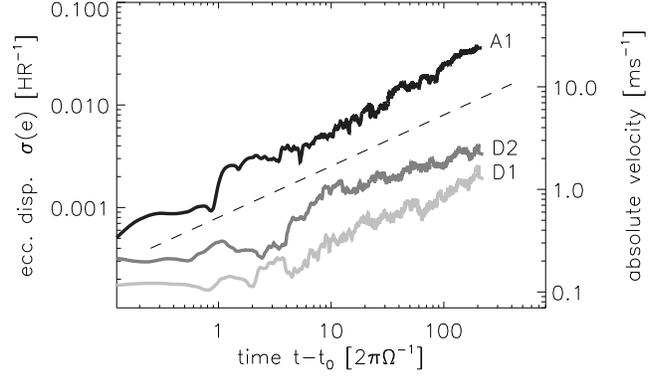}
  \caption{Random-walk eccentricity growth for the dispersion,
    $\sigma(e)$, of sets of 25 particles in each run. The expected
    $\sqrt{t}$ behaviour is indicated by the dashed line. The rhs axis
    gives the corresponding radial velocity dispersion, leading to
    disruptive relative velocities for the fully active run A1. The
    amplitudes for the dead zone runs are lower by a factor of 10-20.}
  \label{fig:vdist_cmp_ecc}
\end{figure}

We now consider the dynamics of larger planetesimals.
The excitation of the eccentricity (and equivalently the radial 
velocity dispersion) can be seen in Fig.~\ref{fig:vdist_cmp_ecc}, 
where we plot the r.m.s. values of these quantities averaged over
the planetesimal swarms as a function of time for the runs A1, D1 and D2.
The time history is consistent with a random-walk behaviour, indicated 
by the dashed line representing an $e(t) \sim \sqrt{t}$ dependence. Expressing
the time evolution of the eccentricity according to
\begin{equation}
e(t)/h = C_{\sigma}(e) \sqrt{t},
\label{eqn:e-v-t}
\end{equation}
where $h=H/r$, and the time is measured in local orbits, we obtain the
following values for the amplitude factors, $C_{\sigma}(e)$ (which are
also listed in Tab.~\ref{tab:sim_results}): $2.68\times10^{-3}$,
$2.50\times10^{-4}$, and $1.54\times10^{-4}$ for models A1, D2, and D1
respectively. Despite the different levels of turbulence (as reflected
in both $\alpha$ and $\Gamma$), the values in
Tab.~\ref{tab:sim_results} show that the stirring amplitude of model
A1 agrees quite well with the unstratified model from paper~I.  The
amplitudes from the dead zone runs D2 and D1 are lower by a factor of
$\simeq11$ and $\simeq17$, respectively, showing that the presence of
the dead zone has a clear and measurable effect on the strength of
eccentricity stirring.

\subsection{Long-term evolution}

As we have only been able to run our simulations for about 200 orbits,
it is important to consider the long term evolution by examining the
magnitude of the equilibrium eccentricity that is expected to arise
through the balance between turbulent eccentricity excitation and gas
drag and/or collisional damping \citep{2008ApJ...686.1292I}.  The
question of whether or not collisional growth of planetesimals may
occur in the different disc models is determined by the velocity
dispersions obtained relative to the disruption/erosion velocity
thresholds, and this point is addressed in the discussion that
follows.

\subsubsection{Gas drag damping}

We have not included gas drag or collisional damping in the
simulations presented in this paper, but gas drag was included in the
runs presented in paper~I, and they showed that over run times of a
few hundred orbits the gas drag has little influence on the
eccentricity for planetesimals with radii $R_p \ge 100\m$.  We adopt
the usual expression for the gas drag force
\citep{1977MNRAS.180...57W}
\begin{equation}
{\bf F}_{\rm drag} = \frac{1}{2} C_{\rm D} \pi R_{\rm p}^2 \rho\
|{\bf v}_{\rm p} - {\bf v}_{\rm g} |\ 
({\bf v}_{\rm p} - {\bf v}_{\rm g})
\label{eqn:drag}
\end{equation}
where $C_{\rm D}$ is the drag coefficient. For making simple estimates
we take the value $C_{\rm D}=0.44$, appropriate for larger
planetesimals for which the Reynolds number of the flow around the
body ${\cal R}_e > 800$.  The eccentricity growth over time is given
by (\ref{eqn:e-v-t}) where the time, $t$, is measured in local
orbits. Following the same procedure adopted in paper~I, and
calculating the equilibrium velocity dispersion by equating the
eccentricity growth time scale and the gas drag-induced damping time
scale, we obtain
\begin{equation}
v_{\rm disp} = \left( \frac{4 C_{\sigma}^2(e) h^2 R_{\rm p} 
\rho_{\rm p} v^2_{\rm k} }{10^9 C_{\rm D} \rho} \right)^{1/3}
\label{eqn:vdisp-drag}
\end{equation}
where $R_{\rm p}$ and $\rho_{\rm p}$ are the planetesimal radius and
density, and $v_{\rm k}$ is the Keplerian velocity. Note that the form
taken by (\ref{eqn:vdisp-drag}) now assumes that the system evolution
time given in (\ref{eqn:e-v-t}) is expressed in seconds.  The
equilibrium velocity dispersions obtained using (\ref{eqn:vdisp-drag})
for each disc model and planetesimal size are listed in the third
column of Tab.~\ref{tab:results}.

\subsubsection{Collisional damping}

The effects of collisional damping, due to inelastic collisions between
planetesimals, may be estimated by noting that the collisional damping
time scale is given by the product of the collision time scale and
the number of collisions required to damp out the kinetic energy of
random motion for a typical planetesimal. This approach is similar
to that adopted by \citet{2008ApJ...686.1292I}. The collision time 
(neglecting the effect of gravitational focusing) is given by
\begin{equation}
t_{\rm coll} = \frac{1}{n_{\rm p} \pi R_{\rm p}^2 v_{\rm disp}},
\label{eqn:t-coll}
\end{equation}
where $n_{\rm p}$ is the number density of planetesimals. This
may be approximated by $n_{\rm p} = \Sigma_{\rm p}/(2\, H_{\rm p}\, m_{\rm p})$,
where $H_{\rm p} = v_{\rm disp}/ \Omega_k$ and $\Sigma_{\rm p}$
is the mass surface density of planetesimals. The coefficient of restitution
is given by $C_{\rm R} = v_{\rm f} / v_{\rm i}$, where $v_{\rm i}$ and
$v_{\rm f}$ are the initial and final velocities associated with a collision.
The number of collisions required to damp $v_{\rm disp}$ is therefore
\begin{equation}
N_{\rm coll} = \frac{1}{1- C_{\rm R}}\,.
\label{eqn:n-coll}
\end{equation}
Combining Equations~(\ref{eqn:t-coll}) and (\ref{eqn:n-coll}) we
obtain the collision damping time:
\begin{equation}
\tau_{\rm c-damp} = \frac{8 R_{\rm p} \rho_{\rm p}}
                         {3 \Sigma_{\rm p} \Omega_{\rm k}}
                    \left(\frac{1}{1- C_{\rm R}} \right)\,.
\label{t-c-damp}
\end{equation}
Equating the collisional damping time with the eccentricity growth
time associated with (\ref{eqn:e-v-t}) -- and given explicitly by
equation (16) in paper~I -- yields an expression for the equilibrium
velocity dispersion
\begin{equation}
v_{\rm disp} = \sqrt{\frac{4 R_{\rm p} \rho_{\rm p} C_{\sigma}^2(e) h^2 v_k^2}
                          {10^9 \Sigma_{\rm p} \Omega_k} 
                     \left(\frac{1}{1-C_{\rm R}}\right)}\,.
\label{eqn:vdisp-coll}
\end{equation}
The value to be adopted for the coefficient of restitution, $C_{\rm
R}$, depends on the material composition, the impact velocity, and
whether or not the planetesimals may be considered to be monolithic
bodies or rubble piles held together by self-gravity alone. Adopting
the velocity dependent formula from \citet{1984Natur.309..333B}
\begin{equation}
C_{\rm R} = {\rm MIN} \, \left\{\ 0.34 \left( \frac{v_{\rm disp}}
{1\,{\rm cm \, s}^{-1}} \right)^{-0.234},\ 1\ \right\}\,,
\label{eqn:C_R}
\end{equation}
we find that $C_{\rm R} \simeq 0$ for the values of the radial
velocity dispersion, $v_{\rm disp}$, that we obtain in the
simulations.  We therefore adopt $C_{\rm R}=0$ when estimating the
equilibrium $v_{\rm disp}$. The equilibrium velocity dispersions
obtained using (\ref{eqn:vdisp-coll}) for each disc model and
planetesimal size are listed in the fourth column of
Table~\ref{tab:results}. In obtaining these values we assume that
$\Sigma_{\rm p} = 4 \times (\Sigma / 250)$ -- i.e. the surface density
of solids is a factor of $1/250$ lower than the gas surface density,
but is augmented by a factor of 4 beyond the ice-line. We further
assume that all disc solids are incorporated within planetesimals of
size $R_{\rm p}$ for each size that we consider.

\subsubsection{Disruption velocity thresholds}

The consequences of the equilibrium velocity dispersions obtained from
equations~(\ref{eqn:vdisp-drag}) and (\ref{eqn:vdisp-coll}) for
planetary growth can only be determined by comparing them with the
disruption/erosion thresholds for colliding bodies
\citep{1999Icar..142....5B,2009ApJ...691L.133S}, and with the escape
velocities associated with the planetesimals. In a recent study,
\citet{2009ApJ...691L.133S} present a universal law for collision
outcomes in the form
\begin{equation}
\frac{M_{\rm lr}}{M_{tot}} = 1 - \frac{1}{2} \frac{Q_{\rm R}}{Q_{\rm D}},
\label{eqn:m-lr}
\end{equation}
where $M_{\rm lr}$ is the mass of the largest post-collision remnant,
$M_{\rm tot}$ is the total mass of the colliding objects $M_1 + M_2$,
and $Q_{\rm R}$ is the reduced mass kinetic energy normalised by
the total mass
\begin{equation}
Q_{\rm R} = \frac{M_1 M_2}{2 M^2_{\rm tot}} V_{\rm I}^2.
\label{eqn:Q-R}
\end{equation}
Here $V_{\rm I}$ is the impact velocity.
For accretion to occur during a collision between equal sized
bodies, we require $M_{\rm lr}/M_{\rm tot} > 1/2$, or
equivalently $Q_{\rm R}/Q_{\rm D} < 1$, such that $Q_{\rm D}$
is the collisional disruption or erosion threshold.

The value of $Q_{\rm D}$ is sensitive to factors that influence
the energy and momentum coupling between colliding bodies
(e.g. impact velocity, strength and porosity). \citet{2009ApJ...691L.133S}
fit results from their numerical simulations and data in the literature
using the expression
\begin{equation}
  Q_{\rm D} = \left[ q_{\rm s} R_{12}^{9 \mu_{\rm m} / (3 - 2 \phi_{\rm m})} 
                   + q_{\rm G} R_{12}^{3 \mu_{\rm m}} \right] 
                   V_{\rm I}^{(2 - 3 \mu_{\rm m})}\,,
\label{eqn:Q-D}
\end{equation}
where $q_{\rm s}$, $q_{\rm G}$, $\phi_{\rm m}$ and $\mu_{\rm m}$ are
parameters.  $R_{12}$ is the spherical radius of the combined mass,
assuming $\rho_{\rm p}=1\g\perccm$. Since we use $\rho_{\rm p} =
2\g\perccm$, and consider collisions between equal-sized bodies only,
we have $R_{12}= 2^{2/3} R_{\rm p}$. The first term on the right hand
side of (\ref{eqn:Q-D}) represents the strength regime, while the
second term represents the gravity regime. Small bodies are supported
by material strength, which decreases as the planetesimal size
increases. On the other hand, gravity increases in importance with
growing planetesimal size -- with the transition between the strength
and gravity regimes occurring at $R_{\rm p} \approx
100\m$. \citet{2009ApJ...691L.133S} derive the following values for
the above parameters for weak aggregates (weak rock): $q_{\rm s} =
500$, $q_{\rm G} =10^{-4}$, $\mu_{\rm m}=0.4$, $\phi_{\rm m}=7$.  For
strong rocks they use basalt laboratory data and modelling results
from \citet{1999Icar..142....5B} to obtain: $q_{\rm s} = 7\times
10^4$, $q_{\rm G} =10^{-4}$, $\mu_{\rm m}=0.5$, $\phi_{\rm m}=8$.  In
the limit of large planetesimals, the disruption curve can be best fit
by their results for colliding rubble piles, for which $Q_{\rm D} =
1.7 \times 10^{-6} R_{12}^2$. Given the uncertainties associated with
the structure and material strength of planetesimals, we tabulate the
disruption velocities, $v_{\rm crit}$, for both weak and strong
planetesimals in the fifth and sixth columns of
Tab.~\ref{tab:results}. For $R_{\rm p} \ge 10\km$, we tabulate the
disruption velocities for rubble piles in the seventh column.

\subsection{Model A1}
\label{sec:A1}

We now consider the results for the fully turbulent model A1.  The
equilibrium velocity dispersions corresponding to gas drag damping are
listed in the third column of Tab.~\ref{tab:results} for $100\m$,
$1\km$, $10\km$ and $100\km$-sized planetesimals (assuming a density
$\rho_{\rm p} =2$ g cm$^{-3}$).  These values of $v_{\rm disp}$ are
smaller than those presented in paper~I by approximately a factor of
two, because in that paper we adopted a slightly more massive disc
model, a larger planetesimal density $\rho_{\rm p}=3$ g cm$^{-3}$, and
we utilised cylindrical disc models that give rise to enhanced
stirring of planetesimals relative to the full 3D simulations
considered here.

Comparing the values of $v_{\rm disp}$ in the third and fourth
columns, we see that for all planetesimal sizes damping due to gas
drag dominates over collisional damping, so it is the values in the
third column that are closest to the true equilibrium values obtained
when all sources of damping act simultaneously. We see that $v_{\rm
disp} > v_{\rm crit}$ for planetesimals composed of both `weak' and
`strong' rock for $R_{\rm p} \le 10\km$, and for $R_{\rm p} = 100\km$
we see that $v_{\rm disp}$ exceeds the disruption/erosion threshold
for rubble piles. We conclude that if planetesimals reach their
equilibrium values for $v_{\rm disp}$, then mutual collisions will
lead to their destruction.

Estimated times for $v_{\rm disp}$ to grow to the equilibrium values
via a random-walk ($v_{\rm disp} \sim \sqrt{t}$) are listed in the
eighth column of Tab.~\ref{tab:results}. The most favourable models
for the incremental formation of planetesimals at $5\au$ in a laminar
disc suggest formation times of a few $\times 10^4\yr$
\citep{2000SSRv...92..295W}. Formation times in a turbulent disc
exceed this because the vertical settling of solids is reduced
\citep{2008A&A...480..859B}. The random velocity growth times in
Tab.~\ref{tab:results} thus indicate that if planetesimals were able
to form incrementally in the disc that we simulate, then they would
always have a velocity dispersion equal to the equilibrium value.
But, the time to reach the disruption velocity $v_{\rm crit}=15.2 \ms$
for strong $1\km$ aggregates is only $800\yr$, much shorter than the
formation time. It is clear that planetesimals cannot form and grow by
a process of collisional agglomeration in a fully turbulent disc
similar to model A1.

\subsection{Models D1 and D2}
\label{sec:D1D2}

We first discuss model D1, which has the deeper dead zone of the two
dead zone simulations. The values of $v_{\rm disp}$ listed in
Tab.~\ref{tab:results} show that for planetesimals with $R_{\rm p} \le
10 \km$ collisional damping dominates gas drag in setting the
equilibrium velocity dispersion. Only for $R_{\rm p}=100\km$ is gas
drag more important. The transition from gas drag dominated damping in
model A1 to collisional damping in model D1 occurs because of the
different functional dependencies on the eccentricity excitation
coefficient, $C_{\sigma}(e)$ in equations~(\ref{eqn:vdisp-drag}) and
(\ref{eqn:vdisp-coll}).

The equilibrium $v_{\rm disp}$ for all planetesimal sizes lie between
the disruption/erosion thresholds, $v_{\rm crit}$, for weak and strong
aggregates.  For the larger planetesimals with $R_{\rm p}=10$ and
$100\km$ we see that $v_{\rm disp} < v_{\rm crit}$ for rubble
piles. These results suggest that collisions between planetesimals in
a dead zone can lead to growth rather than destruction, with the
outcome depending on the material strength.

The times required for $v_{\rm disp}$ to reach the equilibrium values
are listed in the eighth column of Tab.~\ref{tab:results}. Given that
the time required for incremental formation of km-sized planetesimals
in a disc with modest turbulence is likely to exceed $10^5\yr$ at
$5\au$, the time scales for the growth of $v_{\rm disp}$ suggest that
planetesimals growing through a process of particle sticking will
always have velocity dispersions close to the equilibrium values. The
fact that these values are below the disruption thresholds for strong
aggregates suggests that collisional growth may still be possible,
albeit at a slower rate than in a laminar disc.

If km-sized planetesimals can form, further accretion is normally
expected to arise via runaway growth, leading to the formation of
$\sim 1000\km$ sized oligarchs within $\sim 10^5\yr$
\citep{1993Icar..106..190W, 2009ApJ...690L.140K}. Runaway growth
requires the velocity dispersion to be lower than the escape velocity
from the accreting bodies, and for an internal density $\rho_{\rm p}
=2$ g cm$^{-3}$ this is given by $v_{\rm esc} = 10\left(\frac{R_{\rm
p}}{10\km} \right)\ms$. A typical $10\km$ planetesimal in model D1
forming via incremental growth will be excited to $v_{\rm disp} >
10\ms$ during its formation, so turbulent stirring will prevent
runaway growth from occurring for bodies of this size. Because $v_{\rm
disp}$ remains below the disruption threshold for rubble piles,
however, collisions can still lead to growth, but at a substantially
slower rate than occurs during runaway growth. We discuss the
implications of our results for the more rapid planetesimal formation
scenarios presented by \citet{2008ApJ...687.1432C} and
\citet{2007Natur.448.1022J} in Section~\ref{sec:discussion}.

Similar conclusions may be drawn from the results of model D2 as we
have drawn for model D1. But, having a deeper dead zone, as in model
D1, is clearly preferable for planetesimal formation due to the fact
we find that the velocity dispersion is smaller in that case.

\begin{figure}
  \center\includegraphics[width=0.95\columnwidth]{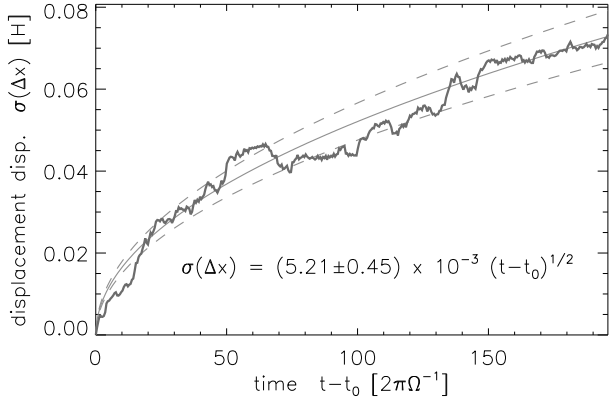}
  \caption{Dispersion of the radial displacement $\Delta x$ of a swarm
    of 25 test particles for run A1. The growing spread in the
    particles' separation is well approximated by a random-walk as
    illustrated by the over-plotted curve.}
  \label{fig:vdist_dx}
\end{figure}

\section{Radial diffusion of planetesimals}
\label{sec:da-grow}

It was shown in paper~I that the radial drift of planetesimals of size
$R_{\rm p} > 10\km$ over typical disc life times due to gas drag is
small. Instead, the diffusion of planetesimal semimajor axes is
dominated by the stochastic gravitational torques exerted on the
planetesimals by the turbulent disc.  This implies that material of
common origin and composition (e.g. planetesimals which form at a
particular radial location in the disc), in time will spread out over
a certain region in semimajor axis.  The degree of spreading is
determined by the disc life time and the strength of the stochastic
torques.

In paper~I, we examined the degree of spreading induced by the
stochastic torques in a fully turbulent disc model similar in mass to
the minimum mass solar nebula model, and showed that over a putative
disc life time of $5\Myr$, planetesimals embedded in the current
asteroid-belt region would spread inward and outward over a distance
of $\sim 2.5\au$. Such a conclusion sits uncomfortably with the
relatively low volatile content of the inner solar system planets
\citep{2006Icar..184...39O}, which would presumably be higher if
substantial volatile-rich material from beyond the snow-line had
migrated inward during the planet-forming epoch. Additionally,
although the picture of how different taxonomic classes of asteroids
are distributed as a function of heliocentric distance has become more
complex in recent years \citep{2003Icar..162...10M}, the earlier work
of \citet{1982Sci...216.1405G} shows a clear trend of increasing
volatile content as a function of heliocentric distance. In this
latter study, the distribution of taxonomic types is consistent with a
picture of {\it in situ} formation, followed by radial diffusion over
a distance of $\simeq 0.5\au$. In the following discussion we judge
that our models have had a measure of success if they do not violate
the constraints implied by this picture.

In Fig.~\ref{fig:vdist_dx}, we plot the evolution of the r.m.s. of the
radial displacement of a swarm of test particles, and we find the
resulting curve to be consistent with a random-walk. As mentioned
above, we neglect dynamical interactions amongst the particles. This
implies that self-stirring due to close encounters is not accounted
for, and the given values have to be regarded as lower limits.

In Sect.~\ref{sec:trq}, we have demonstrated that the gravitational
torques acting on particles can be represented by a normal
distribution, and their temporal correlations possess a finite
coherence time. These two properties allow us to estimate particle
diffusion based on a Fokker-Planck equation. As discussed in detail in
section \textsection3.7 of paper~I \citep[see
also][]{2006ApJ...647.1413J}, the natural variable to describe this
process is the specific angular momentum $j$. The time scale for
diffusion of particle angular momenta due to stochastic torques is
\begin{equation}
t_j = \frac{ (\Delta j)^2}{D_j}
\label{eqn:tdiff}
\end{equation}
where $D_j$ is the diffusion coefficient and $\Delta j$ is the
change in $j$. The diffusion coefficient $D_j$ may be approximated
by $D_j = \sigma_{\Gamma}^2 \tau_{\rm c}$, where $\sigma_{\Gamma}$
is the r.m.s. of the fluctuating specific torques, and $\tau_{\rm c}$ 
is the correlation time associated with the fluctuating torques.

The change in specific angular momentum obtained after an evolution
time of $t$ is given by
\begin{equation}
\Delta j = \sqrt{D_j \, t}.
\label{eqn:delta-j}
\end{equation}
Noting that small changes in specific angular momentum are
related to small changes in the semi-major axis, $a$, by the expression
\begin{equation}
\frac{\Delta j}{j} = \frac{\Delta a }{2 a}
\label{eqn:delta-aj}
\end{equation}
the change in semimajor axis over an evolution time $t$ is:
\begin{equation}
\Delta a = \frac{2 a \sqrt{D_j t}}{j}.
\label{eqn:delta-a}
\end{equation}

\begin{figure}
  \center\includegraphics[width=\columnwidth]{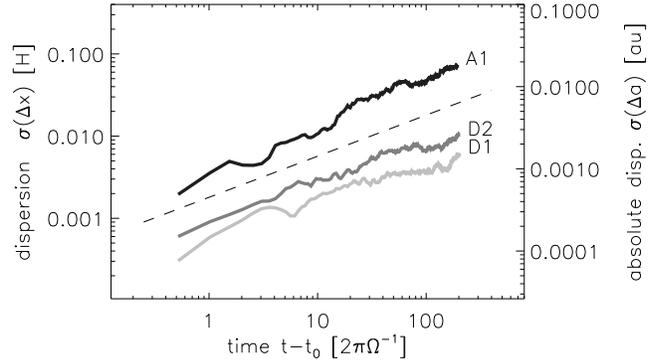}
  \caption{Model comparison for the random-walk behaviour shown in
    Fig.~\ref{fig:vdist_dx} above, but plotted logarithmically. The
    second axis gives the absolute dispersion $\Delta a$ at
    $r=5\au$. Evolution scaling with the square-root of time
    is indicated by the dashed line. }
  \label{fig:vdist_cmp_dx}
\end{figure}

We can now examine the degree of agreement between the level of
particle diffusion observed directly in the simulations, and presented
in Fig.~\ref{fig:vdist_cmp_dx}, and predictions based on
(\ref{eqn:delta-a}). The radial location of the shearing box in our
simulations is assumed to be $5\au$, and simulation run times are $t
\simeq 200$ local orbits. The value of $D_j$ for each model may be
computed using the expression $D_j = \sigma_{\Gamma}^2 \tau_{\rm c}$,
and values for $\sigma_{\Gamma}$, expressed in c.g.s. units, may be
read off Fig.~\ref{fig:trq_his} for each model. The corresponding
estimates of $\tau_{\rm c}$ are listed in Tab.~\ref{tab:sim_results} 
(along with the r.m.s. specific torques, $\sigma_{\Gamma}$).

We note that the time evolution of the r.m.s. radial displacement
obtained in each model, $\sigma_{\Delta x}$, has been fitted using 
the expression 
\begin{equation}
\sigma_{\Delta x}/H = C_{\sigma}(\Delta x) \sqrt{t}
\label{eqn:delta-x-fit}
\end{equation}
where $H$ is the local disc thickness, the time $t$ is measured in
units of local orbits, and the coefficients $C_{\sigma}(\Delta x)$ are
tabulated in Tab.~\ref{tab:sim_results}. After 200 orbits, the radial
diffusion given by (\ref{eqn:delta-x-fit}) corresponds to a change of
semimajor axis $\Delta a = 0.0184\au$ for model A1, which can be
verified by inspection of Figure~\ref{fig:vdist_cmp_dx}. The predicted
change from (\ref{eqn:delta-a}) is $\Delta a = 0.0122\au$,
approximately 30\% smaller than the observed value.

For model D1, we observe a change $\Delta a = 0.00167\au$, and predict
a change from (\ref{eqn:delta-a}) of $\Delta a = 0.00161\au$, giving
excellent agreement. For model D2, the observed change in semimajor
axis is $\Delta a = 0.00256\au$, which is approximately 25\% smaller
than the predicted value of $\Delta a = 0.00336\au$. Taking these
results overall, and noting that the use of only 25 particles in the
simulations implies minimal sampling errors at the 20\% level, we
consider the agreement between the observed and predicted levels of
diffusion to be very satisfactory.

Examining the longer term evolution, we use (\ref{eqn:delta-x-fit}) to
calculate the level of diffusion expected over an assumed disc life
time of $5\Myr$. For the fully active model, A1, we obtain $\Delta a =
2.9\au$, which is in good agreement with the result obtained in
paper~I. For model D1, we obtain $\Delta a = 0.26\au$, and for model
D2 we obtain $\Delta a = 0.40\au$.  It is clear from these values that
in a vertically stratified disc sustaining MHD turbulence without a
dead zone, radial diffusion is predicted to be too large to be
consistent with solar system constraints. For a disc with a
significant dead zone whose height extends either side of the midplane
by a distance $\sim 2 H$, however, we see that the degree of radial
mixing is strongly diminished, and the results are consistent with
the distribution of asteroidal taxonomic types \citep{1982Sci...216.1405G}.


\section{Discussion}
\label{sec:discussion}

We now attempt to summarise our results and present a coherent picture
of how the dynamics and growth of planetesimals are affected by
turbulent stirring in discs with and without dead zones. We frame our
discussion in the context of the three planetesimal formation models
discussed in the introduction: incremental growth through particle
sticking over time scales exceeding $10^5\yr$
\citep{2000SSRv...92..295W, 2008A&A...480..859B}; concentration of
chondrules in turbulent eddies, followed by gravitational contraction
on the chondrule settling time -- typically $\sim 10^2$ - $10^3$
orbits \citep{2008ApJ...687.1432C}; gravitational collapse of dense
clumps of metre-sized bodies formed in turbulent discs through a
combination of trapping in local pressure maxima and the streaming
instability \citep{2007Natur.448.1022J}. Although there are
significant problems with the incremental growth picture, in this
paper we are interested in exploring the effects of turbulence on
macroscopic bodies of sizes $R_{\rm p} \ge 100\m$, and so it is useful
for our discussion to assume an optimistic outcome for this model.
Toward the end of this section we also discuss some of the
shortcomings of our model, and issues that are raised by these for the
interpretation of our results.

\subsection{Turbulent discs without dead zones}

The results presented in Section~\ref{sec:e-growth} suggest that
planetesimal formation and growth via a process involving mutual
collisions between smaller bodies is not possible in fully turbulent
discs. The rapid excitation of large random velocities which exceed
the disruption/erosion threshold for planetesimals with $R_{\rm p} \le
100\km$ will simply lead to the destruction of bodies which form in
this manner. Although the turbulence simulated in this paper is quite
vigorous, because of the imposed vertical magnetic field, the scaling
developed in paper~I for the strength of stochastic forcing as a
function of the effective viscous stress suggest that even an order of
magnitude decrease in the effective viscous $\alpha$ value will not
decrease the random velocities sufficiently to prevent catastrophic
disruption from occurring.

The formation of large ($R_{\rm p} \simeq 100 \km$) planetesimals
through chondrules concentrating in turbulent eddies may be possible
in fully turbulent discs driven by the MRI. The obvious requirement
for there to be a turbulent cascade resulting in Kolmogorov-scale
eddies in which chondrules can concentrate would appear to be
satisfied in such a disc.  During the initial formation and settling
stage of these objects, they are likely to be of low density and hence
strongly coupled to the gas via gas drag. Relative velocities on local
scales relevant for collisions are likely to be small. As they
contract to form ``sandpile" planetesimals, however, they will
decouple from the gas and evolve dynamically like planetesimals with
internal densities of $\rho_{\rm p} \simeq 2\g\perccm$. If
formation/contraction times for these bodies are $\simeq 10^4\yr$ at
$5\au$, then turbulent stirring will cause their velocity dispersion
to grow to the surface escape velocity of $\simeq 100\ms$ within $3.5
\times 10^4\yr$, preventing runaway growth from ensuing.  Further
collisional growth between sandpile planetesimals will therefore occur
slowly. Continued driving of the velocity dispersion by turbulence
eventually allows $v_{\rm disp}$ to exceed the disruption value for
$100\km$ rubble piles within approximately $1.2\Myr$. Clearly
collisional growth must allow significantly larger bodies to form
within this time to prevent the eventual collisional destruction of
the majority of sandpile planetesimals. Even if larger bodies do form
that are safe against collisional disruption, questions remain about
the possibility of forming large planetary cores that can accrete gas
within a few Myr in the absence of runaway growth. A model in which a
relatively small number of large planetesimals avoid collisional
destruction and grow by accreting the surrounding collisional debris
may, however, provide a viable route to the growth of massive
planetary cores.

Rapid formation of large planetesimals through the clumping of
metre-size boulders, followed by their gravitational collapse, has
been demonstrated in fully turbulent discs driven by the MRI
\citep{2007Natur.448.1022J, 2011A&A...529A..62J}.  Although the size
distribution of planetesimals arising from this process is not known
accurately because of numerical limitations, the simulations produce
objects with sizes similar to Ceres ($R_{\rm p} \simeq 500 \km$). Such
objects have surface escape velocities of $500\ms$, and the time over
which turbulent stirring generates a velocity dispersion of this
magnitude is $\sim 8 \times 10^5\yr$. It is unclear whether a
population of planetesimals with an approximately unimodal size
distribution centred on $R_{\rm p}=500 \km$ can undergo runaway
growth, because of self-stirring and weak gas drag damping of
eccentricities.  But if it is feasible then turbulent stirring in a
fully active disc such as computed in model A1 will probably not
provide significant hindrance because of the long time scales required
for the velocity dispersion to grow above $v_{\rm disp} \simeq 500
\ms$.

\subsection{Turbulent discs with dead zones}

Our results for model D1 show that in a model with a relatively deep
dead zone, the equilibrium velocity dispersion is determined by
collisional rather than gas drag damping for bodies with $R_{\rm p}
\le 10\km$, for reasons discussed in section~\ref{sec:D1D2}.  The
equilibrium velocity dispersion lies between the disruption/erosion
thresholds for weak and strong aggregates, suggesting that incremental
collisional growth of planetesimals is possible, but depends on the
material strength of the bodies. For $100\m$ bodies, the time required
to excite the velocity dispersion to the equilibrium value of $4.5\ms$
is $\simeq 2.2 \times 10^4\yr$, comfortably shorter than the
formation time of km-sized planetesimals at $5\au$. Similarly, the
time required to excite $1\km$ bodies to their equilibrium random
velocities is $\simeq 2.2 \times 10^5\yr$, comparable to our assumed
formation time for these bodies through incremental growth.  At each
stage during the incremental growth of planetesimals within the dead
zone, the planetesimals will have velocity dispersions close to the
equilibrium values, which then control in-part the formation and
growth rates. But provided the material strength of the planetesimals
is sufficient (larger than that for weak aggregates), planetesimal
growth should be possible in a dead zone through collisional
agglomeration, so long as the bouncing and metre-size barriers can be
overcome \citep{2008A&A...480..859B, 2010A&A...513A..57Z}.

Runaway growth of km-sized planetesimals requires the velocity
dispersion to be significantly lower than the escape velocity from the
accreting bodies, and for $10\km$ planetesimals $v_{\rm esc}=10 \ms$.
Results from model D1 indicate that a velocity dispersion of this
magnitude will be excited within $\simeq 10^5\yr$, effectively
quenching the runaway growth.  Continued collisional growth can still
proceed, since the velocity dispersion remains below the
disruption/erosion threshold (at least for strong aggregates and
rubble piles), and the equilibrium velocity dispersion for $10\km$
objects $v_{\rm disp} = 51\ms$ is not reached until after $2.2\Myr$.
If growth ensues quickly enough, the eventual formation of bodies with
$R_{\rm p} \simeq 100\km$ will allow runway growth of these
planetesimals to occur.  Model D1 shows that the time required to
raise the velocity dispersion up to the escape velocity $v_{\rm
esc}=100\ms$ for these bodies is $10\Myr$. In this general picture,
the influence of stochastic gravitational forces acting on
planetesimals in a dead zone can cause the postponement of runaway
growth for $10\km$ planetesimals, but once $100\km$ sized bodies form,
rapid runaway growth produce numerous $\sim 1000\km$ sized oligarchs
which will then undergo oligarchic growth to form planetary embryos
and cores.  This general picture of delayed runaway growth is similar
in concept to that suggested by \citet{2001Sci...293.1127K} for planet
formation in binary systems.

The formation of sandpile planetesimals via the concentration of
chondrules within turbulent eddies should be possible in a dead zone,
provided that the Reynolds stress generated there is indicative of a
turbulent flow with an energy cascade resulting in Kolmogorov-eddies
in which millimetre particles can become trapped. At the present time
it is not clear that this is an accurate description of the dead zone
flow, as the Reynolds stress appears to be largely due to spiral
density waves that are stochastically excited in the disc active
regions, and undergo non linear damping as their wavelengths shorten
during radial propagation through the disc
\citep{2009MNRAS.397...64H}.  A detailed analysis of high resolution
simulations is required to examine the nature of the flow in the dead
zone. Assuming that $100\km$ sandpile planetesimals can form, the slow
growth of the velocity dispersion means that turbulent stirring cannot
provide an impediment to subsequent runaway growth.

Similar comments apply to the rapid formation of planetesimals in dead
zones via the mechanism described in \citet{2007Natur.448.1022J}.
Formation in this case is driven to a large degree by trapping of
metre-size bodies in pressure maxima (localised regions of
anticyclonic vorticity) generated by turbulence
\citep{2011A&A...529A..62J}, and these are not as prominent in the
midplane of a disc with a dead zone. This suggests that the streaming
instability models in laminar discs presented in
\citet{2009ApJ...704L..75J} may provide a more promising avenue for
rapid planetesimal formation in dead zones. Once large planetesimals
form, the possible onset of runaway growth will not be inhibited by
the weak stochastic excitation of the velocity dispersion in a dead
zone.

\subsection{Caveats}

We now consider the possible effects of assumptions we
have made in our models that may affect their outcomes and the
conclusions we have drawn from them.

\subsubsection{Disc mass}

The disc model we adopt in this study is lower in mass by a factor of
$\sim 3$ than disc models that have been used as the basis for
successful models of giant planet formation
\citep{1996Icar..124...62P}. But it should be noted that these models
often neglected the effects of planetary migration which can enable
planetary embryos to grow beyond their isolation masses, thus
assisting in the formation of giant planets within expected disc life
times \citep{2005A&A...434..343A}.  If relative density perturbations
in the dead zone remain the same when the disc mass is increased, then
a higher disc mass will lead to a corresponding increase in turbulent
stirring, making planetesimal destruction/erosion more likely. But, in
a disc where the ionisation sources are external (stellar X-rays,
galactic cosmic rays), the column density and mass of the active zone
are almost independent of the total disc mass/surface density, and
depend primarily on the penetration depth of the ionising
sources. Consequently, we may expect the relative density fluctuations
near the midplane in the dead zone to be weaker in a heavier disc than
in a lighter disc, potentially providing a less destructive source of
gravitational stirring. At present it is unclear how turbulent
stirring in a dead zone scales with the disc mass, and studying this
issue is difficult from the computational point of view. The shearing
box models we present here are extended in the $x$ and $y$ directions
compared to those used in most studies, and include $\pm 5.5$ scale
heights above and below the midplane. Increasing the disc mass moves
the boundary between the dead and active zones further from the
midplane, such that the vertical computational domain must be extended
significantly to accommodate the active zone, increasing the
computational expense.  It is our intention to address this issue in a
future study.

\subsubsection{Initial magnetic fields}

We have assumed a particular configuration for the initial
magnetic fields in our simulations that includes a small
but non-negligible net-flux vertical component. It is well
known that the existence and strength of net-flux magnetic
fields changes the saturation level of developed MHD turbulence
in simulations of the MRI \citep{1995ApJ...440..742H}. A 
future study of planetesimal dynamics embedded in turbulent
discs with dead zones should examine the role played by
the field topology and strength.

\subsubsection{Planetesimal size distribution}

In our estimates of collisional damping we assume that all
planetesimals are the same size, and that all solids are incorporated
into planetesimals.  The latter assumption is broadly consistent with
our disc model, in which we assume that 90\% of the dust has been
depleted due to the growth of grains into larger bodies. But the
assumption of a single planetesimal size is not realistic. A more
accurate calculation of the role of collisional damping would require
a self-consistent model of accretion and fragmentation such that the
size distribution is modelled self-consistently.  Unfortunately such a
model goes beyond the scope of this paper.

\subsubsection{Planetesimal scale height}

When estimating the scale height of the planetesimal disc during the
calculation of collisional damping rates, we assumed that $H_{\rm p} =
v_{\rm disp}/\Omega_{\rm k}$ (i.e. the velocity dispersion is
isotropic). In a turbulent disc, however, where a significant
component of the stochastic gravitational field is generated by spiral
density waves that have little vertical structure, we should expect
that the eccentricity growth exceeds that of the inclination. Our
simulation results, however, show that for models D1 and D2 the
inclination growth actually exceeds that of the eccentricity. This
surprising result arises because the disc midplane (centre of mass in
the $z$-direction), oscillates vertically with an amplitude that
reaches $\sim 0.02 H$, causing an oscillation of similar magnitude in
the planetesimals. Future simulations of planetesimals within dead
zones will be computed using global models to examine how robust this
phenomenon is, although we note that a similar vertical oscillation
has been reported in the shearing box simulations presented by 
\citet{2007ApJ...670..805O}.

\subsection{Equation of state}

Wave propagation in discs is strongly influenced by the equation of
state \citep{1990ApJ...364..326L, 1995MNRAS.272..618K}. This may
affect the ability of waves that are excited in the active regions to
propagate into the dead zone, and thus affect the turbulent stirring
there. The simulations presented in this paper all adopted an
isothermal equation of state, whereas protoplanetary discs are
expected to be optically thick at $5\au$.

For parameters typical of T Tauri stars, the primary heating mechanism
near the midplane is the absorption of the radiation emitted by the
disc atmosphere, which in turn heats mainly by absorbing light from
the central star \citep{1997ApJ...490..368C}. The accretion heating
also occurs in the atmosphere, leading to a thoroughly isothermal
interior when a dead zone is present \citep{2011arXiv1104.0004H}. The
role of the equation of state in determining the level of turbulent
stirring at the midplane will be explored in a future publication.


\section{Conclusions}
\label{sec:conclusion}

In this paper we have presented the results of simulations of
planetesimals embedded in local models of protoplanetary discs. The
main aim of this work is to understand the dynamics of planetesimals
in turbulent discs, and to examine how dead zones of
varying vertical thicknesses change the stochastic forcing of the
radial velocity dispersion, and the diffusion of planetesimal
semimajor axes. We consider the influence of gas drag and collisional
damping in counteracting the growth of random velocities due to
turbulent stirring, and estimate the equilibrium velocity dispersions
which result.  The main conclusions of this study are as follows.

The presence and depth of a dead zone have negligible effect on the
measured coherence time of the particle stirring, but only affect its
amplitude as measured through the width of the torque distribution.

In fully turbulent discs without dead zones, we find that the velocity
dispersion of embedded planetesimals grows rapidly, and quickly
exceeds the threshold for disruption for planetesimals in the size
range $100\m \le R_{\rm p} \le 10\km$.  We conclude that planetesimal
formation via collisional accretion of smaller bodies cannot occur in
globally turbulent discs. The direct formation of large $100\km$-sized
planetesimals via turbulent concentration and gravitational
instability \citep{2007Natur.448.1022J,2008ApJ...687.1432C} may occur,
but the excitation of their random velocities may prevent runaway
growth, making it difficult to form large planetary embryos and cores
within reasonable disc life times.

Radial diffusion of planetesimals occurs over distances of $\sim
2.5\au$ in fully turbulent discs over typical disc life times of
$5\Myr$. This is inconsistent with the observed correlation between
asteroid taxonomic types and heliocentric distance in the asteroid
belt, which indicate radial mixing over distances $\le 0.5\au$ during
the formation of the solar system.  It thus seems implausible that the
solar nebula sustained levels of turbulence observed in numerical
simulations of fully turbulent discs.

The stochastic forcing of planetesimal random motions is much weaker
in discs with dead zones. In a dead zone with vertical extent $\pm 2H$
above and below the midplane, the stochastic excitation of planetesimal 
eccentricities is reduced by a factor $\sim 17$ relative to a fully
active disc. The damping of planetesimal random motions in this model
is dominated by collisional damping, and the equilibrium velocity
dispersions lie between the disruption thresholds for weak and strong
aggregates. It thus appears that the collisional growth and formation of
km-sized planetesimals may be possible in a dead zone. But, we note that
the tidal break-up of comet Shoemaker-Levy 9 \citep{1996Icar..121..225A},
and analysis of the collision ejecta from comet Tempel 1 during the Deep 
Impact mission \citep{2007Icar..190..357R}, suggest that cometary bodies
possess modest material strength. It therefore remains uncertain
what range of collision velocities can be tolerated during the incremental
growth of km-sized planetesimals as envisaged in the models of
\citet{2000SSRv...92..295W} and \citet{2008A&A...480..859B}.

Although km-sized planetesimals may be able form within a dead zone,
subject to overcoming the metre-sized and bouncing barriers,
stochastic forcing by the turbulent disc excites random velocities
above the escape velocity for $R_{\rm p}=10\km$ bodies within
$10^5\yr$, quenching runaway growth. Slower growth of planetesimals
through collisions remains possible, and runaway growth can ensue once
bodies with $R_{\rm p} \simeq 100\km$ form, due to the slow growth of
random velocities within a dead zone. If $R_{\rm p} \simeq 1000\km$
`oligarchs' form, oligarchic growth can proceed with very modest
influence from the weak stochastic forcing in the dead zone. A
corollary of this is that $100\km$-sized planetesimals that form
through the concentration of particles and subsequent gravitational
instability in a dead zone will not be prevented from undergoing
runaway growth by turbulent stirring.

In summary, we come to the following specific conclusions regarding
the outcome of our present study:
\begin{itemize}

\item{Comparing the results obtained for dead zones with different
vertical heights ($\sim 1.5H$ and $2H$), we find that the larger dead
zone model clearly favours the collisional formation of planetesimals
because of weaker stochastic forcing of random motions.}

\item{The radial diffusion of planetesimals is much reduced in discs
with dead zones relative to fully active discs. Diffusion over a
distance equal to $0.26\au$ is predicted by our deeper dead zone
model, consistent with solar system constraints.}

\item{The low diffusion levels for planetesimal semimajor axes in dead
zones clearly means that stochastic torques in such an environment
cannot prevent the large scale inward drift of low mass planets via
type I migration.}

\item{We suggest that dead zones provide safe havens for km-sized 
planetesimals against destructive collisions, and are probably a 
required ingredient for the formation of planetary systems.}

\end{itemize}

This paper is part of series in which we are examining the dynamics of
planetesimals and planets in increasingly realistic disc models, and
shows that different qualitative outcomes for planet formation can be
expected in discs with dead zones versus those without. In future
publications we will examine a number of issues, including how changing
the disc surface density and the mass ratio between the active and dead zones 
modifies the turbulent stirring, and how turbulence varies as a function
of heliocentric distance in global MHD simulations of protoplanetary discs.


\section*{Acknowledgements}

This work used the \NIII code developed by Udo Ziegler at the
Astrophysical Institute Potsdam. All computations were performed on
the QMUL HPC facility, purchased under the SRIF initiative. R.P.N. and
O.G. acknowledge the hospitality of the Isaac Newton Institute for
Mathematical Sciences, where part of the work presented in this paper
was completed during the `Dynamics of Discs and Planets' research
programme. N.J.T. was supported by the Jet Propulsion Laboratory,
California Institute of Technology, the NASA Origins and Outer Planets
programs, and the Alexander von Humboldt Foundation. We thank the
referee, Stuart Weidenschilling, for useful comments that led to
improvements to this paper.


\begin{thebibliography}{}

\bibitem[\protect\citeauthoryear{{Alibert}, {Mordasini}, {Benz} \&
  {Winisdoerffer}}{{Alibert} et~al.}{2005}]{2005A&A...434..343A}
{Alibert} Y.,  {Mordasini} C.,  {Benz} W.,    {Winisdoerffer} C.,  2005, \aap,
  434, 343

\bibitem[\protect\citeauthoryear{{Armitage}}{{Armitage}}{1998}]{1998ApJ...501L%
.189A}
{Armitage} P.~J.,  1998, \apjl, 501, L189+

\bibitem[\protect\citeauthoryear{{Asphaug} \& {Benz}}{{Asphaug} \&
  {Benz}}{1996}]{1996Icar..121..225A}
{Asphaug} E.,  {Benz} W.,  1996, \icarus, 121, 225

\bibitem[\protect\citeauthoryear{{Balbus} \& {Hawley}}{{Balbus} \&
  {Hawley}}{1991}]{1991ApJ...376..214B}
{Balbus} S.~A.,  {Hawley} J.~F.,  1991, \apj, 376, 214

\bibitem[\protect\citeauthoryear{{Balsara} \& {Meyer}}{{Balsara} \&
  {Meyer}}{2010}]{2010arXiv1003.0018B}
{Balsara} D.~S.,  {Meyer} C.,  2010, (astro-ph:1003.0018)

\bibitem[\protect\citeauthoryear{{Baruteau} \& {Lin}}{{Baruteau} \&
  {Lin}}{2010}]{2010ApJ...709..759B}
{Baruteau} C.,  {Lin} D.~N.~C.,  2010, \apj, 709, 759

\bibitem[\protect\citeauthoryear{{Benz} \& {Asphaug}}{{Benz} \&
  {Asphaug}}{1999}]{1999Icar..142....5B}
{Benz} W.,  {Asphaug} E.,  1999, Icarus, 142, 5

\bibitem[\protect\citeauthoryear{{Blaes} \& {Balbus}}{{Blaes} \&
  {Balbus}}{1994}]{1994ApJ...421..163B}
{Blaes} O.~M.,  {Balbus} S.~A.,  1994, \apj, 421, 163

\bibitem[\protect\citeauthoryear{{Brandenburg}}{{Brandenburg}}{2008}]{2008AN..%
..329..725B}
{Brandenburg} A.,  2008, \rm AN, 329, 725

\bibitem[\protect\citeauthoryear{{Brandenburg}, {Candelaresi} \&
  {Chatterjee}}{{Brandenburg} et~al.}{2009}]{2009MNRAS.398.1414B}
{Brandenburg} A.,  {Candelaresi} S.,    {Chatterjee} P.,  2009, \mnras, 398,
  1414

\bibitem[\protect\citeauthoryear{{Brauer}, {Dullemond} \& {Henning}}{{Brauer}
  et~al.}{2008}]{2008A&A...480..859B}
{Brauer} F.,  {Dullemond} C.~P.,    {Henning} T.,  2008, \aap, 480, 859

\bibitem[\protect\citeauthoryear{{Bridges}, {Hatzes} \& {Lin}}{{Bridges}
  et~al.}{1984}]{1984Natur.309..333B}
{Bridges} F.~G.,  {Hatzes} A.,    {Lin} D.~N.~C.,  1984, \nat, 309, 333

\bibitem[\protect\citeauthoryear{{Chiang} \& {Goldreich}}{{Chiang} \&
  {Goldreich}}{1997}]{1997ApJ...490..368C}
{Chiang} E.~I.,  {Goldreich} P.,  1997, \apj, 490, 368

\bibitem[\protect\citeauthoryear{{Courvoisier}, {Hughes} \&
  {Proctor}}{{Courvoisier} et~al.}{2010}]{2010AN....331..667C}
{Courvoisier} A.,  {Hughes} D.~W.,    {Proctor} M.~R.~E.,  2010, AN, 331, 667

\bibitem[\protect\citeauthoryear{{Cuzzi}, {Hogan} \& {Shariff}}{{Cuzzi}
  et~al.}{2008}]{2008ApJ...687.1432C}
{Cuzzi} J.~N.,  {Hogan} R.~C.,    {Shariff} K.,  2008, \apj, 687, 1432

\bibitem[\protect\citeauthoryear{{Davis}, {Stone} \& {Pessah}}{{Davis}
  et~al.}{2010}]{2010ApJ...713...52D}
{Davis} S.~W.,  {Stone} J.~M.,    {Pessah} M.~E.,  2010, \apj, 713, 52

\bibitem[\protect\citeauthoryear{{Flaig}, {Kley} \& {Kissmann}}{{Flaig}
  et~al.}{2010}]{2010MNRAS.tmp.1450F}
{Flaig} M.,  {Kley} W.,    {Kissmann} R.,  2010, \mnras, 409, 1297

\bibitem[\protect\citeauthoryear{{Fleming} \& {Stone}}{{Fleming} \&
  {Stone}}{2003}]{2003ApJ...585..908F}
{Fleming} T.,  {Stone} J.~M.,  2003, \apj, 585, 908

\bibitem[\protect\citeauthoryear{{Fleming}, {Stone} \& {Hawley}}{{Fleming}
  et~al.}{2000}]{2000ApJ...530..464F}
{Fleming} T.~P.,  {Stone} J.~M.,    {Hawley} J.~F.,  2000, \apj, 530, 464

\bibitem[\protect\citeauthoryear{{Flock}, {Dzyurkevich}, {Klahr} \&
  {Mignone}}{{Flock} et~al.}{2010}]{2010A&A...516A..26F}
{Flock} M.,  {Dzyurkevich} N.,  {Klahr} H.,    {Mignone} A.,  2010, \aap, 516,
  A26+

\bibitem[\protect\citeauthoryear{{Fromang}, {Terquem} \& {Balbus}}{{Fromang}
  et~al.}{2002}]{2002MNRAS.329...18F}
{Fromang} S.,  {Terquem} C.,    {Balbus} S.~A.,  2002, \mnras, 329, 18

\bibitem[\protect\citeauthoryear{{Gammie}}{{Gammie}}{1996}]{1996ApJ...457..355%
G}
{Gammie} C.~F.,  1996, \apj, 457, 355

\bibitem[\protect\citeauthoryear{{Gardiner} \& {Stone}}{{Gardiner} \&
  {Stone}}{2005}]{2005JCoPh.205..509G}
{Gardiner} T.~A.,  {Stone} J.~M.,  2005, \rm JCoPh, 205, 509

\bibitem[\protect\citeauthoryear{{Garmire}, {Feigelson}, {Broos},
  {Hillenbrand}, {Pravdo}, {Townsley} \& {Tsuboi}}{{Garmire}
  et~al.}{2000}]{2000AJ....120.1426G}
{Garmire} G.,  {Feigelson} E.~D.,  {Broos} P.,  {Hillenbrand} L.~A.,  {Pravdo}
  S.~H.,  {Townsley} L.,    {Tsuboi} Y.,  2000, \aj, 120, 1426

\bibitem[\protect\citeauthoryear{{Glassgold}, {Najita} \& {Igea}}{{Glassgold}
  et~al.}{1997}]{1997ApJ...480..344G}
{Glassgold} A.~E.,  {Najita} J.,    {Igea} J.,  1997, \apj, 480, 344

\bibitem[\protect\citeauthoryear{{Goldreich} \& {Ward}}{{Goldreich} \&
  {Ward}}{1973}]{1973ApJ...183.1051G}
{Goldreich} P.,  {Ward} W.~R.,  1973, \apj, 183, 1051

\bibitem[\protect\citeauthoryear{{Gradie} \& {Tedesco}}{{Gradie} \&
  {Tedesco}}{1982}]{1982Sci...216.1405G}
{Gradie} J.,  {Tedesco} E.,  1982, Science, 216, 1405

\bibitem[\protect\citeauthoryear{{Gressel}}{{Gressel}}{2010}]{2010MNRAS.405...%
41G}
{Gressel} O.,  2010, \mnras, 405, 41

\bibitem[\protect\citeauthoryear{{Gressel} \& {Ziegler}}{{Gressel} \&
  {Ziegler}}{2007}]{2007CoPhC.176..652G}
{Gressel} O.,  {Ziegler} U.,  2007, \rm CoPhC, 176, 652

\bibitem[\protect\citeauthoryear{{Haisch} Jr., {Lada} \& {Lada}}{{Haisch}
  et~al.}{2001}]{2001ApJ...553L.153H}
{Haisch} Jr. K.~E.,  {Lada} E.~A.,    {Lada} C.~J.,  2001, \apjl, 553, L153

\bibitem[\protect\citeauthoryear{{Hanasz}, {Otmianowska-Mazur}, {Kowal} \&
  {Lesch}}{{Hanasz} et~al.}{2009}]{2009A&A...498..335H}
{Hanasz} M.,  {Otmianowska-Mazur} K.,  {Kowal} G.,    {Lesch} H.,  2009, \aap,
  498, 335

\bibitem[\protect\citeauthoryear{{Hawley}}{{Hawley}}{2001}]{2001ApJ...554..534%
H}
{Hawley} J.~F.,  2001, \apj, 554, 534

\bibitem[\protect\citeauthoryear{{Hawley}, {Gammie} \& {Balbus}}{{Hawley}
  et~al.}{1995}]{1995ApJ...440..742H}
{Hawley} J.~F.,  {Gammie} C.~F.,    {Balbus} S.~A.,  1995, \apj, 440, 742

\bibitem[\protect\citeauthoryear{{Hayashi}}{{Hayashi}}{1981}]{1981PThPS..70...%
35H}
{Hayashi} C.,  1981, \rm Progress of Th. Phys. Suppl., 70, 35

\bibitem[\protect\citeauthoryear{{Heinemann} \& {Papaloizou}}{{Heinemann} \&
  {Papaloizou}}{2009a}]{2009MNRAS.397...52H}
{Heinemann} T.,  {Papaloizou} J.~C.~B.,  2009a, \mnras, 397, 52

\bibitem[\protect\citeauthoryear{{Heinemann} \& {Papaloizou}}{{Heinemann} \&
  {Papaloizou}}{2009b}]{2009MNRAS.397...64H}
{Heinemann} T.,  {Papaloizou} J.~C.~B.,  2009b, \mnras, 397, 64

\bibitem[\protect\citeauthoryear{{Hirose} \& {Turner}}{{Hirose} \&
  {Turner}}{2011}]{2011arXiv1104.0004H}
{Hirose} S.,  {Turner} N.~J.,  2011, \apj, 732, L30

\bibitem[\protect\citeauthoryear{{Ida}, {Guillot} \& {Morbidelli}}{{Ida}
  et~al.}{2008}]{2008ApJ...686.1292I}
{Ida} S.,  {Guillot} T.,    {Morbidelli} A.,  2008, \apj, 686, 1292

\bibitem[\protect\citeauthoryear{{Ida} \& {Makino}}{{Ida} \&
  {Makino}}{1993}]{1993Icar..106..210I}
{Ida} S.,  {Makino} J.,  1993, Icarus, 106, 210

\bibitem[\protect\citeauthoryear{{Igea} \& {Glassgold}}{{Igea} \&
  {Glassgold}}{1999}]{1999ApJ...518..848I}
{Igea} J.,  {Glassgold} A.~E.,  1999, \apj, 518, 848

\bibitem[\protect\citeauthoryear{{Ilgner} \& {Nelson}}{{Ilgner} \&
  {Nelson}}{2006a}]{2006A&A...445..205I}
{Ilgner} M.,  {Nelson} R.~P.,  2006a, \aap, 445, 205

\bibitem[\protect\citeauthoryear{{Ilgner} \& {Nelson}}{{Ilgner} \&
  {Nelson}}{2006b}]{2006A&A...445..223I}
{Ilgner} M.,  {Nelson} R.~P.,  2006b, \aap, 445, 223

\bibitem[\protect\citeauthoryear{{Ilgner} \& {Nelson}}{{Ilgner} \&
  {Nelson}}{2008}]{2008A&A...483..815I}
{Ilgner} M.,  {Nelson} R.~P.,  2008, \aap, 483, 815

\bibitem[\protect\citeauthoryear{{Johansen}, {Klahr} \& {Henning}}{{Johansen}
  et~al.}{2011}]{2011A&A...529A..62J}
{Johansen} A.,  {Klahr} H.,    {Henning} T.,  2011, \aap, 529, A62+

\bibitem[\protect\citeauthoryear{{Johansen} \& {Levin}}{{Johansen} \&
  {Levin}}{2008}]{2008A&A...490..501J}
{Johansen} A.,  {Levin} Y.,  2008, \aap, 490, 501

\bibitem[\protect\citeauthoryear{{Johansen}, {Oishi}, {Low}, {Klahr}, {Henning}
  \& {Youdin}}{{Johansen} et~al.}{2007}]{2007Natur.448.1022J}
{Johansen} A.,  {Oishi} J.~S.,  {Low} M.,  {Klahr} H.,  {Henning} T.,
  {Youdin} A.,  2007, \nat, 448, 1022

\bibitem[\protect\citeauthoryear{{Johansen}, {Youdin} \& {Mac Low}}{{Johansen}
  et~al.}{2009}]{2009ApJ...704L..75J}
{Johansen} A.,  {Youdin} A.,    {Mac Low} M.,  2009, \apjl, 704, L75

\bibitem[\protect\citeauthoryear{{Johnson}, {Goodman} \& {Menou}}{{Johnson}
  et~al.}{2006}]{2006ApJ...647.1413J}
{Johnson} E.~T.,  {Goodman} J.,    {Menou} K.,  2006, \apj, 647, 1413

\bibitem[\protect\citeauthoryear{{Kenyon} \& {Bromley}}{{Kenyon} \&
  {Bromley}}{2009}]{2009ApJ...690L.140K}
{Kenyon} S.~J.,  {Bromley} B.~C.,  2009, \apjl, 690, L140

\bibitem[\protect\citeauthoryear{{Kokubo} \& {Ida}}{{Kokubo} \&
  {Ida}}{1998}]{1998Icar..131..171K}
{Kokubo} E.,  {Ida} S.,  1998, Icarus, 131, 171

\bibitem[\protect\citeauthoryear{{Kortenkamp}, {Wetherill} \&
  {Inaba}}{{Kortenkamp} et~al.}{2001}]{2001Sci...293.1127K}
{Kortenkamp} S.~J.,  {Wetherill} G.~W.,    {Inaba} S.,  2001, Science, 293,
  1127

\bibitem[\protect\citeauthoryear{{Korycansky} \& {Pringle}}{{Korycansky} \&
  {Pringle}}{1995}]{1995MNRAS.272..618K}
{Korycansky} D.~G.,  {Pringle} J.~E.,  1995, \mnras, 272, 618

\bibitem[\protect\citeauthoryear{{Latter}, {Fromang} \& {Gressel}}{{Latter}
  et~al.}{2010}]{2010MNRAS.406..848L}
{Latter} H.~N.,  {Fromang} S.,    {Gressel} O.,  2010, \mnras, 406, 848

\bibitem[\protect\citeauthoryear{{Latter}, {Lesaffre} \& {Balbus}}{{Latter}
  et~al.}{2009}]{2009MNRAS.394..715L}
{Latter} H.~N.,  {Lesaffre} P.,    {Balbus} S.~A.,  2009, \mnras, 394, 715

\bibitem[\protect\citeauthoryear{{Lin}, {Papaloizou} \& {Savonije}}{{Lin}
  et~al.}{1990}]{1990ApJ...364..326L}
{Lin} D.~N.~C.,  {Papaloizou} J.~C.~B.,    {Savonije} G.~J.,  1990, \apj, 364,
  326

\bibitem[\protect\citeauthoryear{{Miller} \& {Stone}}{{Miller} \&
  {Stone}}{2000}]{2000ApJ...534..398M}
{Miller} K.~A.,  {Stone} J.~M.,  2000, \apj, 534, 398

\bibitem[\protect\citeauthoryear{{Miyoshi} \& {Kusano}}{{Miyoshi} \&
  {Kusano}}{2005}]{2005JCoPh.208..315M}
{Miyoshi} T.,  {Kusano} K.,  2005, \rm JCoPh, 208, 315

\bibitem[\protect\citeauthoryear{{Moth{\'e}-Diniz}, {Carvano} \&
  {Lazzaro}}{{Moth{\'e}-Diniz} et~al.}{2003}]{2003Icar..162...10M}
{Moth{\'e}-Diniz} T.,  {Carvano} J.~M.~{\'A}.,    {Lazzaro} D.,  2003, \icarus,
  162, 10

\bibitem[\protect\citeauthoryear{{Nelson}}{{Nelson}}{2005}]{2005A&A...443.1067%
N}
{Nelson} R.~P.,  2005, \aap, 443, 1067

\bibitem[\protect\citeauthoryear{{Nelson} \& {Gressel}}{{Nelson} \&
  {Gressel}}{2010}]{2010MNRAS.409..639N}
{Nelson} R.~P.,  {Gressel} O.,  2010, \mnras, 409, 639

\bibitem[\protect\citeauthoryear{{Nelson} \& {Papaloizou}}{{Nelson} \&
  {Papaloizou}}{2004}]{2004MNRAS.350..849N}
{Nelson} R.~P.,  {Papaloizou} J.~C.~B.,  2004, \mnras, 350, 849

\bibitem[\protect\citeauthoryear{{O'Brien}, {Morbidelli} \&
  {Levison}}{{O'Brien} et~al.}{2006}]{2006Icar..184...39O}
{O'Brien} D.~P.,  {Morbidelli} A.,    {Levison} H.~F.,  2006, Icarus, 184, 39

\bibitem[\protect\citeauthoryear{{Oishi}, {Mac Low} \& {Menou}}{{Oishi}
  et~al.}{2007}]{2007ApJ...670..805O}
{Oishi} J.~S.,  {Mac Low} M.,    {Menou} K.,  2007, \apj, 670, 805

\bibitem[\protect\citeauthoryear{{Papaloizou} \& {Nelson}}{{Papaloizou} \&
  {Nelson}}{2003}]{2003MNRAS.339..983P}
{Papaloizou} J.~C.~B.,  {Nelson} R.~P.,  2003, \mnras, 339, 983

\bibitem[\protect\citeauthoryear{{Pessah} \& {Goodman}}{{Pessah} \&
  {Goodman}}{2009}]{2009ApJ...698L..72P}
{Pessah} M.~E.,  {Goodman} J.,  2009, \apjl, 698, L72

\bibitem[\protect\citeauthoryear{{Pollack}, {Hubickyj}, {Bodenheimer},
  {Lissauer}, {Podolak} \& {Greenzweig}}{{Pollack}
  et~al.}{1996}]{1996Icar..124...62P}
{Pollack} J.~B.,  {Hubickyj} O.,  {Bodenheimer} P.,  {Lissauer} J.~J.,
  {Podolak} M.,    {Greenzweig} Y.,  1996, \icarus, 124, 62

\bibitem[\protect\citeauthoryear{{Regev} \& {Umurhan}}{{Regev} \&
  {Umurhan}}{2008}]{2008A&A...481...21R}
{Regev} O.,  {Umurhan} O.~M.,  2008, \aap, 481, 21

\bibitem[\protect\citeauthoryear{{Rheinhardt} \& {Brandenburg}}{{Rheinhardt} \&
  {Brandenburg}}{2010}]{2010A&A...520A..28R}
{Rheinhardt} M.,  {Brandenburg} A.,  2010, \aap, 520, A28+

\bibitem[\protect\citeauthoryear{{Richardson}, {Melosh}, {Lisse} \&
  {Carcich}}{{Richardson} et~al.}{2007}]{2007Icar..190..357R}
{Richardson} J.~E.,  {Melosh} H.~J.,  {Lisse} C.~M.,    {Carcich} B.,  2007,
  \icarus, 190, 357

\bibitem[\protect\citeauthoryear{{R{\"u}diger} \& {Pipin}}{{R{\"u}diger} \&
  {Pipin}}{2000}]{2000A&A...362..756R}
{R{\"u}diger} G.,  {Pipin} V.~V.,  2000, \aap, 362, 756

\bibitem[\protect\citeauthoryear{{Sano}, {Miyama}, {Umebayashi} \&
  {Nakano}}{{Sano} et~al.}{2000}]{2000ApJ...543..486S}
{Sano} T.,  {Miyama} S.~M.,  {Umebayashi} T.,    {Nakano} T.,  2000, \apj, 543,
  486

\bibitem[\protect\citeauthoryear{{Shi}, {Krolik} \& {Hirose}}{{Shi}
  et~al.}{2010}]{2010ApJ...708.1716S}
{Shi} J.,  {Krolik} J.~H.,    {Hirose} S.,  2010, \apj, 708, 1716

\bibitem[\protect\citeauthoryear{{Sicilia-Aguilar}, {Hartmann}, {Brice{\~n}o},
  {Muzerolle} \& {Calvet}}{{Sicilia-Aguilar}
  et~al.}{2004}]{2004AJ....128..805S}
{Sicilia-Aguilar} A.,  {Hartmann} L.~W.,  {Brice{\~n}o} C.,  {Muzerolle} J.,
  {Calvet} N.,  2004, \aj, 128, 805

\bibitem[\protect\citeauthoryear{{Stewart} \& {Leinhardt}}{{Stewart} \&
  {Leinhardt}}{2009}]{2009ApJ...691L.133S}
{Stewart} S.~T.,  {Leinhardt} Z.~M.,  2009, \apjl, 691, L133

\bibitem[\protect\citeauthoryear{{Stone} \& {Gardiner}}{{Stone} \&
  {Gardiner}}{2010}]{2010ApJS..189..142S}
{Stone} J.~M.,  {Gardiner} T.~A.,  2010, \apjs, 189, 142

\bibitem[\protect\citeauthoryear{{Stone}, {Hawley}, {Gammie} \&
  {Balbus}}{{Stone} et~al.}{1996}]{1996ApJ...463..656S}
{Stone} J.~M.,  {Hawley} J.~F.,  {Gammie} C.~F.,    {Balbus} S.~A.,  1996,
  \apj, 463, 656

\bibitem[\protect\citeauthoryear{{Turner} \& {Drake}}{{Turner} \&
  {Drake}}{2009}]{2009ApJ...703.2152T}
{Turner} N.~J.,  {Drake} J.~F.,  2009, \apj, 703, 2152

\bibitem[\protect\citeauthoryear{{Turner} \& {Sano}}{{Turner} \&
  {Sano}}{2008}]{2008ApJ...679L.131T}
{Turner} N.~J.,  {Sano} T.,  2008, \apjl, 679, L131

\bibitem[\protect\citeauthoryear{{Turner}, {Sano} \& {Dziourkevitch}}{{Turner}
  et~al.}{2007}]{2007ApJ...659..729T}
{Turner} N.~J.,  {Sano} T.,    {Dziourkevitch} N.,  2007, \apj, 659, 729

\bibitem[\protect\citeauthoryear{{Umebayashi}}{{Umebayashi}}{1983}]{1983PThPh.%
.69..480U}
{Umebayashi} T.,  1983, Progress of Theoretical Physics, 69, 480

\bibitem[\protect\citeauthoryear{{Umebayashi} \& {Nakano}}{{Umebayashi} \&
  {Nakano}}{1981}]{1981PASJ...33..617U}
{Umebayashi} T.,  {Nakano} T.,  1981, \pasj, 33, 617

\bibitem[\protect\citeauthoryear{{Umebayashi} \& {Nakano}}{{Umebayashi} \&
  {Nakano}}{2009}]{2009ApJ...690...69U}
{Umebayashi} T.,  {Nakano} T.,  2009, \apj, 690, 69

\bibitem[\protect\citeauthoryear{{Wardle}}{{Wardle}}{1999}]{1999MNRAS.307..849%
W}
{Wardle} M.,  1999, \mnras, 307, 849

\bibitem[\protect\citeauthoryear{{Weidenschilling}}{{Weidenschilling}}{1977}]{%
1977MNRAS.180...57W}
{Weidenschilling} S.~J.,  1977, \mnras, 180, 57

\bibitem[\protect\citeauthoryear{{Weidenschilling}}{{Weidenschilling}}{2000}]{%
2000SSRv...92..295W}
{Weidenschilling} S.~J.,  2000, \ssr, 92, 295

\bibitem[\protect\citeauthoryear{{Weidenschilling} \&
  {Cuzzi}}{{Weidenschilling} \& {Cuzzi}}{1993}]{1993prpl.conf.1031W}
{Weidenschilling} S.~J.,  {Cuzzi} J.~N.,  1993, in {E.~H.~Levy \& J.~I.~Lunine}
  eds., Protostars and Planets III, Univ. Arizona Press, Tucson, p. 1031

\bibitem[\protect\citeauthoryear{{Wetherill} \& {Stewart}}{{Wetherill} \&
  {Stewart}}{1993}]{1993Icar..106..190W}
{Wetherill} G.~W.,  {Stewart} G.~R.,  1993, Icarus, 106, 190

\bibitem[\protect\citeauthoryear{{Yang}, {Mac Low} \& {Menou}}{{Yang}
  et~al.}{2009}]{2009ApJ...707.1233Y}
{Yang} C.,  {Mac Low} M.,    {Menou} K.,  2009, \apj, 707, 1233

\bibitem[\protect\citeauthoryear{{Youdin} \& {Goodman}}{{Youdin} \&
  {Goodman}}{2005}]{2005ApJ...620..459Y}
{Youdin} A.~N.,  {Goodman} J.,  2005, \apj, 620, 459

\bibitem[\protect\citeauthoryear{{Ziegler}}{{Ziegler}}{2004}]{2004JCoPh.196..3%
93Z}
{Ziegler} U.,  2004, \rm JCoPh, 196, 393

\bibitem[\protect\citeauthoryear{{Ziegler}}{{Ziegler}}{2008}]{2008CoPhC.179..2%
27Z}
{Ziegler} U.,  2008, \rm CoPhC, 179, 227

\bibitem[\protect\citeauthoryear{{Zsom}, {Ormel}, {G{\"u}ttler}, {Blum} \&
  {Dullemond}}{{Zsom} et~al.}{2010}]{2010A&A...513A..57Z}
{Zsom} A.,  {Ormel} C.~W.,  {G{\"u}ttler} C.,  {Blum} J.,    {Dullemond} C.~P.,
   2010, \aap, 513, A57+

\end{thebibliography}

\appendix
\bsp

\label{lastpage}

\end{document}